\documentclass[preprint,showpacs,preprintnumbers,amsmath,amssymb,showkeys]{revtex4}

\usepackage{graphicx}
\usepackage{dcolumn}
\usepackage{bm}
\usepackage{braket}
\usepackage{enumitem}
\usepackage[utf8]{inputenc}
\usepackage{amsmath}
\usepackage{amssymb}
\usepackage{amsthm}

\begin{document}

\noindent

\preprint{}

\title{Operational interpretation and estimation of quantum trace-norm asymmetry based on weak value measurement and some bounds}

\author{Agung Budiyono}
\email{agungbymlati@gmail.com}
\affiliation{Research Center for Quantum Physics, National Research and Innovation Agency, South Tangerang 15314, Indonesia} 
\affiliation{Research Center for Nanoscience and Nanotechnology, Bandung Institute of Technology, Bandung, 40132, Indonesia}
\affiliation{Department of Engineering Physics, Bandung Institute of Technology, Bandung, 40132, Indonesia}

\date{\today}

\begin{abstract} 

The asymmetry of a quantum state relative to a translational group is a central concept in many areas of quantum science and technology. An important and geometrically intuitive measure of translational asymmetry of a state is given by the trace-norm asymmetry, which is defined as the trace norm of the commutator between the state and the generator of the translation group. While trace-norm asymmetry satisfies all the requirements for a bonafide measure of translational asymmetry of a state within the quantum resource theoretical framework, its meaning in terms of laboratory operations is still missing. Here, we first show that the trace-norm asymmetry is equal to the average absolute imaginary part of the weak value of the generator of the translation group optimized over all possible orthonormal bases of the Hilbert space. Hence, it can be estimated via the measurement of weak value combined with a classical optimization in the fashion of quantum variational circuit which may be implemented using the near-term quantum hardware. We then use the link between the trace-norm asymmetry and the nonreal weak value to derive the relation between the trace-norm asymmetry with other basic concepts in quantum statistics. We further obtain trade-off relations for the trace-norm asymmetry and quantum Fisher information, having analogous forms to the Kennard-Weyl-Robertson uncertainty relation. 

\end{abstract} 

\maketitle       

\section{Introduction}

Quantum information theory promises novel schemes of information processing technology that are exceedingly far more efficient and secure than those based on classical means. This is achieved by harnessing various nonclassical aspects of quantum mechanics. One such aspect, which has garnered a lot of attention lately, is the concept of quantum asymmetry \cite{Bartlett asymmetry review,Marvian - Spekkens speakable and unspeakable coherence}. Quantum asymmetry captures the behaviour of a quantum state under (the unitary representation of) certain group of transformation. Like other nonclassical aspects of quantum mechanics, it originates from the quantum superposition principle. Significant efforts over the past decade have shown that this manifestation of quantum superposition as asymmetry is a prerequisite for quantum frame alignment \cite{Marvian application of coherence as asymmetry for aligning reference frame,Gour asymmetry - reference frame,Vaccaro asymmetry - reference frame} and quantum metrology \cite{Marvian coherence as asymmetry 0,Marvian coherence as asymmetry}, and it is a key concept in the study of quantum speed limit \cite{Marvian coherence measure quantum speed limit,Mondal asymmetry and speed limit} and quantum thermodynamics \cite{Vaccaro asymmetry - reference frame,Aberg coherence in quantum thermodynamics,Lostaglio coherence in quantum thermodynamics 1,Lostaglio coherence in quantum thermodynamics 2,Cwiklinski coherence in quantum thermodynamics,Yang coherence in quantum thermodynamics}. It is thus important to be able to characterize and quantify the asymmetry of an unknown quantum state using well-defined operations in laboratory. 

Consider a Hermitian operator $K$ on a finite-dimensional Hilbert space generating a one-parameter group of translation unitaries: $\{U_{K,\theta}=e^{-iK\theta},\theta\in\mathbb{R}\}$. A quantum state represented by a density operator $\varrho$ on the Hilbert space is symmetric relative to the translation group generated by $K$ if it is invariant under the translation unitaries, i.e., $e^{-iK\theta}\varrho e^{iK\theta}=\varrho$ for all $\theta\in\mathbb{R}$. All other states are asymmetric relative to the translation group. For a quantum state $\varrho$ to be symmetric relative to the translation group generated by $K$, the state must therefore commute with the generator of the translation, i.e., $[K,\varrho]:=K\varrho-\varrho K=0$, so that they are jointly diagonalizable. Assuming that $K$ is nondegenerate, and denoting the eigenstates of $K$ as $\{\ket{k}\}$, we thus have: $\varrho=\sum_kp_k\ket{k}\bra{k}$, where $\{p_k\}$ are the real and nonnegative eigenvalues of $\varrho$ satisfying $\sum_kp_k=1$. Hence, a symmetric state relative to a translation group is a convex combination or a classical mixture of the eigenstates $\{\ket{k}\}$ of the generator $K$ of the translation. This means that a translationally asymmetric state is a superposition of some elements of the eigenstates $\{\ket{k}\}$ of $K$. Namely, it is coherent with respect to the orthonormal basis $\{\ket{k}\}$ \cite{Marvian - Spekkens speakable and unspeakable coherence}. 

The asymmetry of a quantum state relative to a group of translation is better understood by regarding it as a resource in some information processing tasks. This insight has led to the application of the rigorous mathematical framework of quantum resource theory \cite{Horodecki resource theory - review,Chitambar resource theory - review} to characterize, quantify and manipulate the translational asymmetry \cite{Gour quantifying quantum asymmetry 1,Gour asymmetry - reference frame,Marvian PhD thesis asymmetry and quantum information}. In the general framework of quantum resource theory, permissible quantum operations are restricted to those which are easily implemented reflecting certain physical and/or operational constraints. Such operations are regarded as free. Accordingly, quantum states are divided into those that can be prepared by the set of free operations called free states, and those that cannot, which are regarded as resourceful states. In the resource theory of translational asymmetry, the free operations are given naturally by the set of translationally-covariant quantum operations, i.e., those which commute with the translation unitaries \cite{Marvian - Spekkens speakable and unspeakable coherence}. Such operations cannot create asymmetric states from symmetric states relative to the translation group. Hence, symmetric states are regarded as free, and asymmetric states are resourceful.     

An important measure of translational asymmetry obtained within the framework of quantum resource theory is the trace-norm asymmetry \cite{Marvian - Spekkens speakable and unspeakable coherence,Marvian coherence measure quantum speed limit}. The trace-norm asymmetry of a state $\varrho$ relative to a translation group generated by a Hermitian operator $K$ is defined as
\begin{eqnarray}
A_{\rm Tr}(\varrho;K):=\frac{1}{2}\|[\varrho,K]\|_1,
\label{trace-norm asymmetry}
\end{eqnarray} 
where $\|O\|_1:={\rm Tr}\sqrt{OO^{\dagger}}$ is the trace norm (also called 1-norm) of the operator $O$. The trace-norm asymmetry is not only geometrically intuitive, it also satisfies certain plausible requirements for a bonafide measure of asymmetry within the resource theoretical framework. Most importantly, it satisfies i) faithfulness, i.e., it vanishes if and only if the state is symmetric, and ii) monotonicity, i.e., it is nonincreasing under the translationally-covariant operations: $A_{\rm Tr}(\Phi(\varrho);K)\le A_{\rm Tr}(\varrho;K)$, where $\Phi(\cdot)$ is a completely positive trace-nonincreasing linear map satisfying the translationally-covariant condition: $\Phi(e^{-iK\theta}\varrho e^{iK\theta})=e^{-iK\theta}\Phi(\varrho)e^{iK\theta}$. While the trace-norm asymmetry offers a closed formula, its meaning in terms of laboratory operations is not clear except when the state is pure, $\varrho=\ket{\psi}\bra{\psi}$, in the case of which the trace-norm asymmetry can be expressed as: 
\begin{eqnarray}
A_{\rm Tr}(\ket{\psi}\bra{\psi};K)=\Delta_K(\ket{\psi}\bra{\psi}), 
\label{trace-norm asymmetry for pure state is equal to twice quantum standard deviation}
\end{eqnarray}
where, for a generic state $\varrho$, $\Delta_K(\varrho)^2:={\rm Tr}(K^2\varrho)-({\rm Tr}(K\varrho))^2$ is the quantum variance of the outcomes of the measurement of observable $K$ over the state $\varrho$. A better understanding on the operational meaning of the trace-norm asymmetry may suggest a fresh insight into its application to characterize certain quantum (information) protocols, and its estimation in the laboratory. It may also reveal the relation between the trace-norm asymmetry and other measures of asymmetry and quantum coherence, and between the trace-norm asymmetry and other basic concepts in quantum statistics.

In the present work, we first show that the trace-norm asymmetry relative to a translation group is equal to the average absolute nonreal part of the weak value \cite{Aharonov weak value,Wiseman weak value,Aharonov-Daniel book,Dressel weak value review,Tamir weak value review} of the generator of the translation, maximized over all possible orthonormal bases of the Hilbert space. Hence, it can be estimated in experiment directly, i.e., without recoursing to full state tomography, through the measurement of the weak value \cite{Aharonov weak value,Wiseman weak value,Dressel weak value review,Lundeen complex weak value,Jozsa complex weak value,Johansen quantum state from successive projective measurement,Johansen weak value from a sequence of strong measurement,Lostaglio KD quasiprobability and quantum fluctuation,Haapasalo generalized weak value,Cohen estimating of weak value with strong measurements,Vallone strong measurement to reconstruct quantum wave function,Wagner measuring weak values and KD quasiprobability} combined with a classical optimization procedure, in the fashion of variational quantum circuit \cite{Cerezo VQA review}. These estimation schemes should be realizable using the presently available NISQ (Noisy Intermediate-Scale Quantum) hardware \cite{Preskill NISQ era quantum computing}. Moreover, they lend themselves for the operational interpretation of the trace-norm asymmetry. Using the mathematical link between the trace-norm asymmetry and the nonreal part of the weak value, we then derive upper bounds for the trace-norm asymmetry in terms of quantum standard deviation, quantum Fisher information \cite{Holevo book on quantum statistics,Helstrom estimation-based UR,Braunstein estimation-based UR 1,Braunstein estimation-based UR 2,Paris quantum estimation review}, non-real (nonclassical) values of the Kirkwood-Dirac (KD) quasiprobability \cite{Kirkwood quasiprobability,Dirac quasiprobability,Chaturvedi KD distribution}, $l_1$-norm coherence \cite{Baumgratz quantum coherence measure}, and purity of the quantum state. We also obtain a lower bound for the trace-norm asymmetry and the quantum Fisher information in terms of the maximum average noncommutativity between the generator of the translation group and any other bounded Hermitian operators on the Hilbert space. This leads to the derivation of trade-off relations for the trace-norm asymmetry and the quantum Fisher information similar to the Kennard-Weyl-Robertson uncertainty relation, suggesting an interpretation as the trade-off relation for the genuine quantum part of the uncertainty. Analytical computations of the results for the case of a single qubit is given in the Appendix \ref{Some analytical computations for a single qubit}.

\section{Operational interpretation and estimation of trace-norm asymmetry via weak value measurement}

Let us first summarize the concept of weak value whose statistics we will use to characterize the trace-norm asymmetry defined in Eq. (\ref{trace-norm asymmetry}). \\
{\bf Definition 1}. The weak value associated with a Hermitian operator $K$ on a Hilbert space $\mathcal{H}$ with a preselected state represented by a density operator $\varrho$ on $\mathcal{H}$ and a postselected pure state represented by a ray $\ket{\phi}$ in $\mathcal{H}$ is defined as follows \cite{Aharonov weak value,Aharonov-Daniel book,Dressel weak value review,Wiseman weak value,Tamir weak value review}: 
\begin{eqnarray}
K_{\rm w}(\Pi_{\phi}|\varrho):=\frac{{\rm Tr}(\Pi_{\phi}K\varrho)}{{\rm Tr}(\Pi_{\phi}\varrho)}, 
\label{complex weak value}
\end{eqnarray} 
where $\Pi_{\phi}:=\ket{\phi}\bra{\phi}$ is a projector over a subset of the Hilbert space spanned by $\ket{\phi}$, and we have assumed ${\rm Tr}(\Pi_{\phi}\varrho)\neq 0$. \\
Note that the weak value is in general a complex number. Moreover, its real part may lie outside of the eigenvalues spectrum of $K$. Such complex weak values, and weak values with real part lying outside of the spectrum of $K$, are called strange weak values, and have been used to prove quantum contextuality \cite{Pusey negative TMH quasiprobability and contextuality,Kunjwal KD quasiprobability and contextuality,Lostaglio nonreal KD distribution and contextuality in linear response,Lostaglio TMH quasiprobability fluctuation Proposition contextuality}. In the past decade, there has been a surge of interest in the concept of strange weak values, in particular for its close relation with the anomalous values of the KD quasiprobability \cite{Kirkwood quasiprobability,Dirac quasiprobability,Chaturvedi KD distribution}, in a broad fields of quantum science and technology: quantum state tomography \cite{Lundeen direct measurement of wave function,Lundeen measurement of KD distribution,Maccone comparison between direct state measurement and tomography}, quantum thermodynamics \cite{Lostaglio contextuality in quantum linear response,Allahverdyan TMH as quasiprobability distribution of work,Levy quasiprobability distribution for heat fluctuation in quantum regime}, quantum metrology \cite{Lostaglio contextuality in quantum linear response,Arvidsson-Shukur quantum advantage in postselected metrology,Lupu-Gladstein negativity enhanced quantum phase estimation 2022}, quantum information scrambling or quantum chaos in many body systems \cite{Halpern quasiprobability and information scrambling,Alonso KD quasiprobability witnesses quantum scrambling}, and the characterization of different forms of quantum fluctuations \cite{Lostaglio KD quasiprobability and quantum fluctuation}. It has also been very recently used to characterize coherence and asymmetry \cite{Agung KD-nonreality coherence,Agung translational asymmetry from nonreal weak value}. Remarkably, the real and imaginary parts of the weak value can be measured or estimated in experiment via a number of methods \cite{Aharonov weak value,Wiseman weak value,Dressel weak value review,Lundeen complex weak value,Jozsa complex weak value,Johansen quantum state from successive projective measurement,Johansen weak value from a sequence of strong measurement,Lostaglio KD quasiprobability and quantum fluctuation,Haapasalo generalized weak value,Cohen estimating of weak value with strong measurements,Vallone strong measurement to reconstruct quantum wave function,Wagner measuring weak values and KD quasiprobability}. Below we shall be concerned specifically with the imaginary part of the weak value.  

We show that, combined with a classical optimization procedure, the measurement of weak value can be used to estimate the trace-norm asymmetry of an unknown quantum state in laboratory, without first recoursing to quantum state tomography. Hereon, we shall consider quantum systems with a finite-dimensional Hilbert space. 

First, let us define the following quantity which we introduced earlier in Ref. \cite{Agung translational asymmetry from nonreal weak value}.\\ 
{\bf Definition 2}. Let $\mathcal{B}_o(\mathcal{H})$ denote the set of all the orthonormal bases of a Hilbert space $\mathcal{H}$. Then, given a state $\varrho$ and a Hermitian operator $K$ on $\mathcal{H}$, we define a real-valued nonnegative quantity $A_{\rm w}(\varrho;K)$ as the average of the absolute imaginary part of the weak value $K_{\rm w}(\Pi_x|\varrho)$ defined in Eq. (\ref{complex weak value}) over the probability ${\rm Pr}(x|\varrho)={\rm Tr}(\Pi_x\varrho)$ to get $x$ in the measurement described by a projection-valued measure $\{\Pi_x\}$, maximized over all the orthonormal bases $\{\ket{x}\}$ of the Hilbert space, i.e.: 
\begin{eqnarray}
A_{\rm w}(\varrho;K)&:=&\sup_{\{\ket{x}\}\in\mathcal{B}_o(\mathcal{H})}\sum_x\big|{\rm Im}K_{\rm w}(\Pi_x|\varrho)\big|{\rm Pr}(x|\varrho)\nonumber\\
&=&\sup_{\{\ket{x}\}\in\mathcal{B}_o(\mathcal{H})}\sum_x\big|{\rm Im}\braket{x|K\varrho|x}\big|\nonumber\\ 
&=&\frac{1}{2}\sup_{\{\ket{x}\}\in\mathcal{B}_o(\mathcal{H})}\sum_x\big|\braket{x|[K,\varrho]|x}\big|. 
\label{maximum average imaginary part of weak value and incompatibility}
\end{eqnarray}

It is clear from the last line of Eq. (\ref{maximum average imaginary part of weak value and incompatibility}) that $A_{\rm w}(\varrho;K)$ captures the maximum noncommutativity between the state $\varrho$ and the Hermitian operator $K$ over all possible orthonormal bases $\{\ket{x}\}\in\mathcal{B}_o(\mathcal{H})$ under the $l_1$-norm. We showed in Ref. \cite{Agung translational asymmetry from nonreal weak value} that it can be used to quantify the asymmetry of $\varrho$ relative to the translation group generated by $K$ fulfilling certain plausible requirements. Below, we argue that  $A_{\rm w}(\varrho;K)$ is in fact equal to the trace-norm asymmetry defined in Eq. (\ref{trace-norm asymmetry}). 

{\bf Proposition 1}. The trace-norm asymmetry of a state $\varrho$ relative to a translation group generated by a Hermitian operator $K$ defined in Eq. (\ref{trace-norm asymmetry}) can be expressed in terms of the imaginary part of the weak value of $K$ as 
\begin{eqnarray}
A_{\rm Tr}(\varrho;K)=A_{\rm w}(\varrho;K), 
\label{trace-norm asymmetry is equal to maximum total sum of imaginary part of weak value}
\end{eqnarray}
where $A_{\rm w}(\varrho;K)$ is defined in Eq. (\ref{maximum average imaginary part of weak value and incompatibility}). \\
{\bf Proof.} First, recall that the trace norm of an operator $O$ is given by the total sum of the singular values or the eigenvalues modulus of the operator, i.e., $\|O\|_1=\sum_i|o_i|$, where $\{o_i\}$ is the set of eigenvalues of $O$. Next, note that $[K,\varrho]$ is a skew Hermitian operator. Hence, it has the following spectral decomposition $[K,\varrho]=\sum_i\lambda_i\ket{\lambda_i}\bra{\lambda_i}$, where $\{\lambda_i\}$ is the set of purely imaginary eigenvalues of $[K,\varrho]$ with the corresponding orthonormal set of eigenvectors $\{\ket{\lambda_i}\}$. We thus have, upon inserting this into the right-hand side of Eq. (\ref{maximum average imaginary part of weak value and incompatibility}),  
\begin{eqnarray}
\label{trace norm from maximum imaginary part of weak value step 1}
A_{\rm w}(\varrho;K)&=&\frac{1}{2}\sup_{\{\ket{x}\}\in\mathcal{B}_o(\mathcal{H})}\sum_{x}\big|\sum_i\lambda_i\braket{x|\lambda_i}\braket{\lambda_i|x}\big|\nonumber\\
\label{trace norm from maximum imaginary part of weak value step 2}
&=&\frac{1}{2}\sum_{x_*}\big|\sum_i\lambda_i\braket{x_*|\lambda_i}\braket{\lambda_i|x_*}\big|\\
\label{trace norm from maximum imaginary part of weak value step 3}
&\le&\frac{1}{2}\sum_i|\lambda_i|\sum_{x_*}|\braket{x_*|\lambda_i}|^2\\
\label{trace norm from maximum imaginary part of weak value step 3.5}
&=&\frac{1}{2}\sum_i|\lambda_i|=\frac{1}{2}\|[\varrho,K]\|_1\\
\label{trace norm from maximum imaginary part of weak value step 4}
&=&A_{\rm Tr}(\varrho;K). 
\end{eqnarray}
Here, $\{\ket{x_*}\}$ in Eq. (\ref{trace norm from maximum imaginary part of weak value step 2}) is an orthonormal basis which reaches the supremum, the inequality in Eq. (\ref{trace norm from maximum imaginary part of weak value step 3}) is due to the triangle inequality, and we have used the completeness relation for $\{\ket{x_*}\}$ to obtain Eq. (\ref{trace norm from maximum imaginary part of weak value step 3.5}). On the other hand, one can see in Eq. (\ref{trace norm from maximum imaginary part of weak value step 2}) that the equality, i.e., the upper bound, is always attained by choosing $\{\ket{x_*}\}=\{\ket{\lambda_j}\}$ so that we get Eq. (\ref{trace-norm asymmetry is equal to maximum total sum of imaginary part of weak value}). \qed. 

Proposition 1 thus reveals a link between two seemingly different basic concepts of quantum mechanics: the trace-norm asymmetry which quantifies the amount of asymmetry of the state relative to a translation group, and the anomalous nonreal part of the weak value of the generator of the translation group. A concrete analytical computation of the equality in Eq. (\ref{trace-norm asymmetry is equal to maximum total sum of imaginary part of weak value}) for a single qubit with arbitrary state and generator of translation group is given in the Appendix \ref{Proposition 1 for a single qubit}. Note that the computation of the left-hand side of Eq. (\ref{trace-norm asymmetry is equal to maximum total sum of imaginary part of weak value}) is equivalent to finding a basis of the Hilbert space which diagonalizes $[K,\varrho]$. By contrast, to compute the right-hand side of Eq. (\ref{trace-norm asymmetry is equal to maximum total sum of imaginary part of weak value}), we have to find a basis which optimizes Eq. (\ref{maximum average imaginary part of weak value and incompatibility}).

As mentioned earlier, the weak value can be experimentally obtained via a number of methods \cite{Aharonov weak value,Wiseman weak value,Dressel weak value review,Lundeen complex weak value,Jozsa complex weak value,Johansen quantum state from successive projective measurement,Johansen weak value from a sequence of strong measurement,Lostaglio KD quasiprobability and quantum fluctuation,Haapasalo generalized weak value,Cohen estimating of weak value with strong measurements,Vallone strong measurement to reconstruct quantum wave function,Wagner measuring weak values and KD quasiprobability}. Proposition 1 thus offers a scheme to experimentally estimate the trace-norm asymmetry of an unknown quantum state relative to a translation group, directly, i.e., without recoursing to full state tomograpy. To do this, one first makes the measurement of the weak value $K_{\rm w}(\Pi_x|\varrho)$ of the generator of the translation, averages its absolute imaginary part over the probability of outcomes ${\rm Pr}(x|\varrho)={\rm Tr}(\Pi_x\varrho)$ of projective von Neumann measurement $\{\Pi_x\}$, and maximizes over all possible orthonormal bases $\{\ket{x(\vec{\lambda})}\}\in\mathcal{B}_o(\mathcal{H})$ of the Hilbert space $\mathcal{H}$, where $\vec{\lambda}$ is the parameters whose variation over their ranges of values scans all the orthonormal bases of the Hilbert space. The optimization over $\vec{\lambda}$ is carried out by using some classical methods. This estimation scheme of the trace-norm asymmetry therefore requires the ability to implement a parameterized unitary circuit $V_{\vec{\lambda}}$ which prepares all the orthonormal bases $\{\ket{x(\vec{\lambda})}\}$ of the Hilbert space from the standard basis. Hence, we have a hybrid quantum-classical scheme in the fashion of quantum variational circuit \cite{Cerezo VQA review} which can be realized using the NISQ hardware \cite{Preskill NISQ era quantum computing}. This scheme of estimation of the trace-norm asymmetry thus provides an operational interpretation. In contrast to this, the estimation of the trace-norm asymmetry based on full state tomography followed by diagonalization clearly does not offer operational interpretation of the trace-norm asymmetry.   

The equality in Eq. (\ref{trace-norm asymmetry is equal to maximum total sum of imaginary part of weak value}) suggests that the trace-norm asymmetry can be given physical and statistical interpretation in terms of those of the weak values. For example, within the scheme of the estimation of the weak value based on weak measurement with postselection \cite{Aharonov weak value,Wiseman weak value,Aharonov-Daniel book,Dressel weak value review,Tamir weak value review}, $A_{\rm w}(\varrho;K)$, and thus the trace-norm asymmetry $A_{\rm Tr}(\varrho;K)$ of the state $\varrho$ relative to the translation group generated by $K$, can be interpreted as the maximal disturbance of the state due to the translation unitary generated by $K$ \cite{Dressel imaginary weak value and disturbance,Aharonov imaginary weak value and disturbance}. This goes in line with the fact that the trace-norm asymmetry indeed gives the rate of change of the state under the translation unitary generated by $K$ as $\|U_{K,\delta\theta}\varrho U^{\dagger}_{K,\delta\theta}-\varrho\|_1=\|[\varrho,K]\|_1\delta\theta+o(\delta\theta^2)$. This is also the reason why larger trace-norm asymmetry is desirable in quantum parameter estimation as will be corroborated in the next section. On the other hand, within the scheme of the measurement of weak value based on two sequences of strong measurement \cite{Johansen quantum state from successive projective measurement,Johansen weak value from a sequence of strong measurement}, $A_{\rm w}(\varrho;K)$ can be interpreted as the maximal disturbance of the state $\varrho$ due to the nonselective measurement of $\{\Pi_k\}$, the eigenbasis of $K$. Moreover, following the statistical interpretation of weak value developed in Refs. \cite{Johansen weak value best estimation,Hall prior information,Luo imaginary part of weak value as quantum fluctuation}, $A_{\rm w}(\varrho;K)$ can also be interpreted as the maximum absolute error in the optimal estimation of $K$ based on the outcomes $\{x\}$ of the projective measurement described by $\{\Pi_x\}$. Finally, we note that the idea that the imaginary part of the weak value captures the quantum fluctuations has been put forward in Refs. \cite{Luo imaginary part of weak value as quantum fluctuation,Agung imaginary part of the weak value as quantum fluctuation}.  

\section{Upper and lower bounds for trace-norm asymmetry and uncertainty relations}

As an immediate application of the mathematical equality of Eq. (\ref{trace-norm asymmetry is equal to maximum total sum of imaginary part of weak value}) connecting the trace-norm asymmetry relative to a translation group and the nonreal part of the weak value of the generator of the translation, we may obtain results for the former by studying the statistics of the latter. Using this approach, in this section, we derive some relations between the trace-norm asymmetry and certain important concepts in quantum statistics.  

\subsection{Upper bounds: quantum standard deviation, quantum Fisher information, $l_1$-norm coherence and purity}

First, we have the following proposition. \\
{\bf Proposition 2}. The trace-norm asymmetry of a state $\varrho$ relative to a translation group generated by a Hermitian operator $K$ is bounded from above by the quantum standard deviation of $K$ over $\varrho$, i.e., 
\begin{eqnarray}
A_{\rm Tr}(\varrho;K)&\le&\Delta_K(\varrho), 
\label{trace-norm asymmetry is upper bounded by quantum uncertainty}
\end{eqnarray} 
where equality is reached for all pure states. \\
{\bf Proof}. 
First, as shown in the Appendix \ref{maximum total sum of imaginary part is upper bounded by quantum uncertainty}, $A_{\rm w}(\varrho;K)$ defined in Eq. (\ref{maximum average imaginary part of weak value and incompatibility}) is upper bounded by the quantum standard deviation of $K$ over $\varrho$ as
\begin{eqnarray}
A_{\rm w}(\varrho;K)\le \Delta_K(\varrho).
\label{TC w-coherence is upper bounded by quantum standard deviation}
\end{eqnarray}  
Combining this with Eq. (\ref{trace-norm asymmetry is equal to maximum total sum of imaginary part of weak value}), we thus obtain Eq. (\ref{trace-norm asymmetry is upper bounded by quantum uncertainty}). For pure states, as mentioned in Eq. (\ref{trace-norm asymmetry for pure state is equal to twice quantum standard deviation}), the trace-norm asymmetry of $\varrho=\ket{\psi}\bra{\psi}$ relative to the translation group generated by $K$ is exactly equal to the quantum standard deviation of $K$ over $\varrho=\ket{\psi}\bra{\psi}$ so that the inequality in Eq. (\ref{trace-norm asymmetry is upper bounded by quantum uncertainty}) becomes equality. The case of a single qubit is given in the Appendix \ref{trace-norm asymmetry vs quantum uncertainty for a single qubit}. \qed

Proposition 2 thus generalizes Theorem 1 of Ref. \cite{Agung translational asymmetry from nonreal weak value} where we have derived Eq. (\ref{TC w-coherence is upper bounded by quantum standard deviation}) and showed that equality is obtained for a pure state single qubit.  A different sketch of a proof of Proposition 2 based on state purification and the fact that $A_{\rm Tr}(\varrho;K)=\|[\varrho,K]\|_1/2$ is a monotonic measure of asymmetry and $\Delta_K(\varrho)$ is not, is suggested in Ref. \cite{Marvian coherence measure quantum speed limit}. Here we have proven it using a decomposition of the trace-norm asymmetry in terms of the average absolute imaginary part of the weak value of the generator of the translation relative to which the asymmetry is defined. 

Note further that unlike the trace-norm asymmetry (or any other measures of coherence), the upper bound $\Delta_K(\varrho)$ in Eq. (\ref{trace-norm asymmetry is upper bounded by quantum uncertainty}), i.e., the quantum standard deviation, is not sensitive to whether the state is pure or mixed. To see this, consider for instance two extreme cases of maximally coherent state $\ket{\psi_{\rm mc}}=\frac{1}{\sqrt{d}}\sum_ke^{i\theta_k}\ket{k}$ and maximally mixed state: $\varrho_{\rm mm}=\mathbb{I}/d$. Then, in both cases, the upper bound in Eq. (\ref{trace-norm asymmetry is upper bounded by quantum uncertainty}) yields the same value: $\Delta_K(\ket{\psi_{\rm mc}}\bra{\psi_{\rm mc}})=\Delta_K(\varrho_{\rm mm})$. Hence, quantum standard deviation cannot distinguish the maximally coherent state from the maximally mixed state. This is because the quantum standard deviation does not only capture the genuine quantum uncertainty arising from the noncommutativity between the state $\varrho$ and the generator $K$, it also counts the uncertainty arising from the (classical) statistical mixing in the preparation of the state $\varrho$ when it is not pure. It is desirable to have an upper bound which depends on the purity of the state. Such a bound will be given later. 

Proposition 2 thus suggests that $A_{\rm Tr}(\varrho;K)=A_{\rm w}(\varrho;K)$ can be seen as to capture the genuine quantum part of the uncertainty arising in the measurement of the observable $K$ over the quantum system prepared in a state $\varrho$ which originates from their noncommutativity. Namely, $A_{\rm Tr}(\varrho;K)=A_{\rm w}(\varrho;K)$ satisfies the following plausible requirements for any quantity which quantifies the genuine quantum uncertainty of $K$ in $\varrho$ \cite{Luo's genuine quantum uncertainty1,Luo's genuine quantum uncertainty2,Korzekwa quantum-classical decomposition,Hall quantum-classical decomposition}: (i) it vanishes if and only if $K$ and $\varrho$ commute, (ii) it is convex, i.e., $A_{\rm w}(\sum_jp_j\varrho_j;K)\le\sum_j p_j A_{\rm w}(\varrho_j;K)$ where $\{p_j\}$, $\sum_jp_j=1$, are probabilities of preparing the system in the states $\{\varrho_j\}$, and (iii) it is upper bounded by the quantum standard deviation and they are equal for all pure states. The property (ii) of convexity of $A_{\rm w}(\varrho;K)$ can be seen directly from the definition of $A_{\rm w}(\varrho;K)$ in Eq. (\ref{maximum average imaginary part of weak value and incompatibility}) due to the triangle inequality. The above observation also suggests to interpret the difference between the quantum standard deviation of $K$ over $\varrho$ and the trace-norm asymmetry of $\varrho$ relative to a translation group generated by $K$, i.e., $\Delta_K(\varrho)-A_{\rm w}(\varrho;K)$, as  the classical part of the measurement uncertainty. 

Let us proceed to use the equality in Eq. (\ref{trace-norm asymmetry is equal to maximum total sum of imaginary part of weak value}) to explore a connection between the trace-norm asymmetry and quantum Fisher information \cite{Helstrom estimation-based UR,Holevo book on quantum statistics,Braunstein estimation-based UR 1,Braunstein estimation-based UR 2,Paris quantum estimation review}. Consider an imprinting of a scalar parameter $\theta$ to the quantum state of a probe via a quantum process: $\varrho_{\theta}=\Phi_{\theta}(\varrho)$, where $\Phi_{\theta}$ is a completely positive trace-preserving map, and $\varrho$ is the initial quantum state of the probe. Let $\mathcal{M}_{\rm POVM}(\mathcal{H})$ denote the set of all POVMs (positive operator-valued measure): $\{M_x\}$, $M_x\ge 0$, $\sum_xM_x=\mathbb{I}$, describing the most general measurement allowed by quantum mechanics with the outcomes $\{x\}$, when the post-measurement states are not of concern. \\
{\bf Definition 3}. The quantum Fisher information about the parameter $\theta$ encoded in $\varrho_{\theta}$ is defined as \cite{Braunstein estimation-based UR 1,Braunstein estimation-based UR 2,Paris quantum estimation review}
\begin{eqnarray}
&&\mathcal{J}_{\theta}(\Phi_{\theta}(\varrho))\nonumber\\
&:=&\sup_{\{M_x\}\in\mathcal{M}_{\rm POVM}(\mathcal{H})}\sum_x\big(\partial_\theta\ln{\rm Pr}(x|\varrho_{\theta})\big)^2{\rm Pr}(x|\varrho_{\theta}), 
\label{quantum Fisher information}
\end{eqnarray} 
where ${\rm Pr}(x|\varrho_{\theta})={\rm Tr}(M_x\varrho_{\theta})$ is the probability to get $x$ in the measurement described by a POVM $\{M_x\}$. 

Quantum Fisher information is a central quantity in quantum metrology based on quantum parameter estimation \cite{Helstrom estimation-based UR,Holevo book on quantum statistics,Giovannetti quantum estimation review}, wherein one wishes to estimate the value of the parameter $\theta$ encoded in the quantum state of the probe $\varrho_{\theta}$ via some measurement $\{M_x\}$. It characterizes the optimal precision of such parameter estimation based on the quantum Cram\'er-Rao inequality \cite{Braunstein estimation-based UR 1,Braunstein estimation-based UR 2,Paris quantum estimation review}. Below, we are interested in the case where the state of the probe $\varrho_{\theta}$ is obtained using a translation unitary generated by a Hermitian operator $K$, i.e., $\varrho_{\theta}=\Phi_{\theta}(\varrho)=e^{-iK\theta}\varrho e^{iK\theta}$, and denote the associated quantum Fisher information as $\mathcal{J}_{\theta}(\varrho_{\theta};K)$. It is known that, in this case, the quantum Fisher information is independent of the parameter $\theta$, i.e., $\mathcal{J}_{\theta}(\varrho_{\theta};K)=\mathcal{J}_{\theta}(\varrho;K)$. Moreover, it has also been shown that the quantum Fisher information $\mathcal{J}_{\theta}(\varrho;K)$ is a faithful and monotonic measure of the asymmetry of the state $\varrho$ relative to the translation group generated by $K$ \cite{Marvian - Spekkens speakable and unspeakable coherence}. It is thus instructive to study the relation between the quantum Fisher information and the trace-norm asymmetry. 

We obtain the following result. \\ 
{\bf Proposition 3}. The trace-norm asymmetry of $\varrho$ relative to a translation group generated by a Hermitian operator $K$ is upper bounded by the quantum Fisher information about a parameter $\theta$ contained in the state $\varrho_{\theta}$ obtained via a unitary imprinting generated by $K$ as
\begin{eqnarray}
A_{\rm Tr}(\varrho;K)^2\le \mathcal{J}_{\theta}(\varrho;K)/4. 
\label{trace-norm asymmetry is upper bounded by the quantum Fisher information}
\end{eqnarray} 
Moreover, for pure states, the inequality becomes equality and the supremum in Eq. (\ref{quantum Fisher information}) is obtained by a measurement described by a projection-valued measure. \\
{\bf Proof}. First, from Eq. (\ref{quantum Fisher information}) and noting the fact that the set $\mathcal{M}_{\rm POVM}(\mathcal{H})$ of measurements described by POVM $\{M_x\}$ includes the set $\mathcal{M}_{\rm PVM}(\mathcal{H})$ of measurements described by projection-valued measure $\{\Pi_x\}$, we have
\begin{eqnarray}
&&\mathcal{J}_{\theta}(\varrho;K)=\mathcal{J}_{\theta}(\varrho_{\theta};K)\nonumber\\
&\ge&\sup_{\{\Pi_x\}\in\mathcal{M}_{\rm PVM}(\mathcal{H})}\sum_x\big(\partial_\theta\ln{\rm Pr}(x|\varrho_{\theta})\big)^2{\rm Pr}(x|\varrho_{\theta}).
\label{trace-norm asymmetry is upper bounded by the quantum Fisher information step 1}
\end{eqnarray}
On the other hand, from the unitary imprinting: $\varrho_{\theta}=e^{-iK\theta}\varrho e^{iK\theta}$, we have $\partial_{\theta}\varrho_{\theta}=-i[K,\varrho_{\theta}]$, so that noting ${\rm Pr}(x|\varrho_{\theta})={\rm Tr}(\Pi_x\varrho_{\theta})$, the imaginary part of the weak value of $K$ with the preselected state $\varrho_{\theta}$ and postselected state $\ket{x}$ can be expressed as 
\begin{eqnarray}
{\rm Im}K_{\rm w}(\Pi_x|\varrho_{\theta})=\frac{1}{2i}\frac{\braket{x|[K,\varrho_{\theta}]|x}}{\braket{x|\varrho_{\theta}|x}}=\frac{1}{2}\frac{\partial_{\theta}{\rm Pr}(x|\varrho_{\theta})}{{\rm Pr}(x|\varrho_{\theta})}. 
\label{nonreal weak value vs score function}
\end{eqnarray}
Using this relation in Eq. (\ref{trace-norm asymmetry is upper bounded by the quantum Fisher information step 1}), we thus obtain  
\begin{eqnarray}
&&\mathcal{J}_{\theta}(\varrho;K)\nonumber\\
&\ge&4\sup_{\{\ket{x}\}\in\mathcal{B}_o(\mathcal{H})}\sum_x({\rm Im}K_{\rm w}(\Pi_x|\varrho_{\theta}))^2{\rm Tr}(\Pi_x\varrho_{\theta})\nonumber\\
\label{trace-norm asymmetry is upper bounded by the quantum Fisher information step 3}
&\ge&4\big(\sup_{\{\ket{x}\}\in\mathcal{B}_o(\mathcal{H})}\sum_x|{\rm Im}K_{\rm w}(\Pi_x|\varrho_{\theta})|{\rm Tr}(\Pi_x\varrho_{\theta})\big)^2\\
\label{trace-norm asymmetry is upper bounded by the quantum Fisher information step 4}
&=&4A_{\rm w}(\varrho_{\theta};K)^2=4A_{\rm w}(\varrho;K)^2\\
\label{trace-norm asymmetry is upper bounded by the quantum Fisher information step 5}
&=&4A_{\rm Tr}(\varrho;K)^2. 
\end{eqnarray}
Here, Eq. (\ref{trace-norm asymmetry is upper bounded by the quantum Fisher information step 3}) holds due to the Jensen inequality, and to get Eq. (\ref{trace-norm asymmetry is upper bounded by the quantum Fisher information step 4}) we have used Eq. (\ref{maximum average imaginary part of weak value and incompatibility}) and the fact that $A_{\rm w}(\varrho;K)$ is invariant under translation unitary $U_{K,\theta}=e^{-iK\theta}$, i.e.: $A_{\rm w}(\varrho_{\theta};K)=A_{\rm w}(U_{K,\theta}\varrho U_{K,\theta}^{\dagger};K)=A_{\rm w}(\varrho;K)$ which can be proven directly from the defintion \cite{Agung translational asymmetry from nonreal weak value}. Finally Eq. (\ref{trace-norm asymmetry is upper bounded by the quantum Fisher information step 5}) is just Eq. (\ref{trace-norm asymmetry is equal to maximum total sum of imaginary part of weak value}). 

For pure states, $\varrho=\ket{\psi}\bra{\psi}$, from Proposition 2 we have $A_{\rm Tr}(\ket{\psi}\bra{\psi};K)^2=\Delta_K^2(\ket{\psi}\bra{\psi})$. On the other hand, it is known that for pure states with the unitary imprinting: $\ket{\psi_{\theta}}=e^{-iK\theta}\ket{\psi}$, we also have: $\mathcal{J}_{\theta}(\ket{\psi}\bra{\psi};K)=4\Delta_K^2(\ket{\psi}\bra{\psi})$ \cite{Paris quantum estimation review}. From these two equalities for pure states, we thus obtain Eq. (\ref{trace-norm asymmetry is upper bounded by the quantum Fisher information}) with inequality replaced by equality, i.e., $A_{\rm Tr}(\varrho;K)^2=\mathcal{J}_{\theta}(\varrho;K)/4$. Note that as shown in the Appendix \ref{Trace-norm asymmetry vs Fisher information for a single qubit}, for a single qubit, this equality applies even for arbitary mixed state. Hence, for pure states $\varrho=\ket{\psi}\bra{\psi}$, the quantum Fisher information can be expressed as, using Eqs. (\ref{trace-norm asymmetry is equal to maximum total sum of imaginary part of weak value}) and (\ref{maximum average imaginary part of weak value and incompatibility}):
\begin{eqnarray}
&&\mathcal{J}_{\theta}(\ket{\psi}\bra{\psi};K)\nonumber\\
&=&4A_{\rm Tr}(\varrho;K)^2=4A_{\rm w}(\varrho;K)^2\nonumber\\
&=&4\sup_{\{\ket{x}\}\in\mathcal{B}_o(\mathcal{H})}\big(\sum_x\big|{\rm Im}K_{\rm w}(\Pi_x|\varrho)\big|{\rm Pr}(x|\varrho)\big)^2\nonumber\\
&=&\sup_{\{\Pi_x\}\in\mathcal{M}_{\rm PVM}(\mathcal{H})}\big(\sum_x\Big|\frac{\partial_{\theta}{\rm Pr}(x|\varrho)}{{\rm Pr}(x|\varrho)}\Big|{\rm Pr}(x|\varrho)\big)^2,
\label{Fisher information for pure state is equal trace-norm}
\end{eqnarray}
where we have again used Eq. (\ref{nonreal weak value vs score function}). Comparing Eq. (\ref{Fisher information for pure state is equal trace-norm}) to Eq. (\ref{quantum Fisher information}), for pure states, the supremum in Eq. (\ref{quantum Fisher information}) is thus obtained for measurement described by a projection-valued measure. \qed

Eq. (\ref{trace-norm asymmetry is upper bounded by the quantum Fisher information}) of Proposition 3 in particular shows that a quantum state $\varrho$ with larger trace-norm asymmetry relative to a translation group generated by a Hermitian operator $K$ is sufficient for a larger quantum Fisher information about $\theta$ conjugate to $K$. In view of the quantum Cram\'er-Rao inequality, such a state is thus desirable, i.e., it may lead to a better precision, in quantum parameter estimation of $\theta$.  

Next, we use the result of Eq. (\ref{trace-norm asymmetry is equal to maximum total sum of imaginary part of weak value}) to connect the trace-norm asymmetry to an apparently different concept of nonclassicality captured by the nonclassical values of Kirkwood-Dirac (KD) quasiprobability. \\
{\bf Definition 4}. The KD quasiprobability associated with a state $\varrho$ on a Hilbert space $\mathcal{H}$ over a pair of orthonormal bases $\{\ket{x}\}\in\mathcal{B}_o(\mathcal{H})$ and $\{\ket{k}\}\in\mathcal{B}_o(\mathcal{H})$ is defined as \cite{Kirkwood quasiprobability,Dirac quasiprobability,Chaturvedi KD distribution}
\begin{eqnarray}
{\rm Pr}_{\rm KD}(k,x|\varrho):={\rm Tr}(\Pi_x\Pi_k\varrho).
\label{KD quasiprobability}
\end{eqnarray}
KD quasiprobability gives correct marginal probabilities, i.e., $\sum_x{\rm Pr}_{\rm KD}(k,x|\varrho)={\rm Tr}(\Pi_k\varrho)={\rm Pr}(k|\varrho)$ and $\sum_k{\rm Pr}_{\rm KD}(k,x|\varrho)={\rm Tr}(\Pi_x\varrho)={\rm Pr}(x|\varrho)$, where ${\rm Pr}(\cdot)$ is the classical, i.e., real and nonnegative, probability. However, because of the noncommutativity among the state and the projection valued measures $\{\Pi_x\}$ and $\{\Pi_k\}$ corresponding to the two defining orthonormal bases, unlike the Kolmogorovian classical probability, KD quasiprobability may assume complex values and its real part may be negative. In this sense, the nonreality and/or the negativity of the KD quasiprobability therefore captures a form of nonclassicality. Remarkably, the nonreality and/or the negativity of KD quasiprobability, a.k.a KD nonclassicality, has been shown to be tighter than noncommutativity \cite{Drori nonclassicality tighter and noncommutativity,deBievre nonclassicality in KD distribution}. Moreover, recent works showed that KD nonclassicality plays crucial roles in various areas of quantum science \cite{Pusey negative TMH quasiprobability and contextuality,Kunjwal KD quasiprobability and contextuality,Lostaglio nonreal KD distribution and contextuality in linear response,Lostaglio TMH quasiprobability fluctuation Proposition contextuality,Lundeen direct measurement of wave function,Lundeen measurement of KD distribution,Maccone comparison between direct state measurement and tomography,Lostaglio contextuality in quantum linear response,Allahverdyan TMH as quasiprobability distribution of work,Levy quasiprobability distribution for heat fluctuation in quantum regime,Arvidsson-Shukur quantum advantage in postselected metrology,Lupu-Gladstein negativity enhanced quantum phase estimation 2022,Halpern quasiprobability and information scrambling,Alonso KD quasiprobability witnesses quantum scrambling,Lostaglio KD quasiprobability and quantum fluctuation,Agung KD-nonreality coherence,Agung translational asymmetry from nonreal weak value}. 

We further introduce the following normalized trace-norm asymmetry.\\
{\bf Definition 5}. The normalized trace-norm asymmetry of $\varrho$ relative to the translation group generated by a Hermitian operator $K$ is defined as  
\begin{eqnarray}
\tilde{A}_{\rm Tr}(\varrho;K):=A_{\rm Tr}(\varrho;K)/\|K\|_{\rm max}=A_{\rm Tr}(\varrho;\tilde{K}), 
\label{normalized trace-norm asymmetry}
\end{eqnarray}
where for any bounded Hermitian operator $O$ on finite-dimensional Hilbert space, $\tilde{O}$ is defined as $\tilde{O}:=O/\|O\|_{\rm max}$ with $\|O\|_{\rm max}$ the spectral radius, i.e., the maximum singular value, of $O$. \\

We then have the following result. \\
{\bf Proposition 4}. The normalized trace-norm asymmetry of $\varrho$ relative to the translation group generated by a Hermitian operator $K$ is upper bounded by the total sum of the absolute imaginary part of the KD quasiprobability defined over the eigenbasis $\{\ket{k}\}$ of $K$, and a second orthonormal basis of the Hilbert space, maximized over all possible choices of the latter as 
\begin{eqnarray}
\tilde{A}_{\rm Tr}(\varrho;K)&\le&\sup_{\{\ket{x}\}\in\mathcal{B}_o(\mathcal{H})}\sum_{k,x}\big|{\rm Im}{\rm Pr}_{\rm KD}(k,x|\varrho)\big|\nonumber\\
&:=&C_{\rm KD}(\varrho;\{\ket{k}\}). 
\label{normalized trace-norm asymmetry vs imaginary KD quasiprobability}
\end{eqnarray}
{\bf Proof}. Using Eqs. (\ref{trace-norm asymmetry is equal to maximum total sum of imaginary part of weak value}), (\ref{maximum average imaginary part of weak value and incompatibility}), and (\ref{KD quasiprobability}), we first have the following relation: 
\begin{eqnarray}
A_{\rm Tr}(\varrho;K)&=&\sup_{\{\ket{x}\}\in\mathcal{B}_o(\mathcal{H})}\sum_x\big|\sum_kk{\rm Im}{\rm Tr}(\Pi_x\Pi_k\varrho)\big|\nonumber\\
&\le&\|K\|_{\rm max}\sup_{\{\ket{x}\}\in\mathcal{B}_o(\mathcal{H})}\sum_{k,x}\big|{\rm Im}{\rm Pr}_{\rm KD}(k,x|\varrho)\big|,
\label{trace-norm asymmetry vs imaginary KD quasiprobability 1}
\end{eqnarray}
where we have used the spectral decomposition $K=\sum_kk\ket{k}\bra{k}$. Dividing both sides with the spectral radius of $K$, i.e., $\|K\|_{\rm max}$, and noting Eq. (\ref{normalized trace-norm asymmetry}), we obtain Eq. (\ref{normalized trace-norm asymmetry vs imaginary KD quasiprobability}). For a single qubit, the inequality in Eq. (\ref{normalized trace-norm asymmetry vs imaginary KD quasiprobability}) can be checked analytically, as shown in the Appendix \ref{Trace-norm asymmetry vs nonreality of KD quasiprobability}. \qed

Next, it was shown in Ref. \cite{Agung KD-nonreality coherence} that the right-hand side of Eq. (\ref{normalized trace-norm asymmetry vs imaginary KD quasiprobability}) gives a lower bound to the $l_1$-norm coherence \cite{Baumgratz quantum coherence measure} of the state $\varrho$ relative to the orthonormal basis $\{\ket{k}\}$ defined as $C_{l_1}(\varrho;\{\ket{k}\}):=\sum_{k\neq k'}|\braket{k|\varrho|k'}|$, i.e., 
\begin{eqnarray}
C_{\rm KD}(\varrho;\{\ket{k}\})\le C_{l_1}(\varrho;\{\ket{k}\}). 
\label{KD-nonreality coherence is upper bounded by l1-norm coherence}
\end{eqnarray}
Moreover, the inequality becomes equality for an arbitrary state of a single qubit. Noting this, we thus obtain the first corollary of Proposition 4.\\ 
{\bf Corollary 1}. The normalized trace-norm asymmetry of a state $\varrho$ relative to a translation group generated by a Hermitian operator $K$, is upper bounded by the $l_1$-norm coherence of $\varrho$ relative to the orthonormal basis $\{\ket{k}\}$ of $K$, i.e.,
\begin{eqnarray}
\tilde{A}_{\rm Tr}(\varrho;K)\le C_{l_1}(\varrho;\{\ket{k}\}). 
\label{normalized trace-norm asymmetry vs KD coherence and l1-norm coherence}
\end{eqnarray}
Moreover, for a single qubit, assuming the eigenvalues of $K$ are $\{1,-1\}$, the above inequality becomes an equality.\\ 
{\bf Proof}. First, the inequality is obtained by chaining the inequalities of Eqs. (\ref{normalized trace-norm asymmetry vs imaginary KD quasiprobability}) and (\ref{KD-nonreality coherence is upper bounded by l1-norm coherence}). To prove the second half of the corollary, we note that for the case of a single qubit with $K$ having the spectrum of eigenvalues $\{1,-1\}$, we have, as shown in the Appendix \ref{Proposition 1 for a single qubit} (see Eq. (\ref{trace norm for a single qubit general})), $\tilde{A}_{\rm Tr}(\varrho;K)=2|\braket{k_+|\varrho|k_-}|$, where $\ket{k_{\pm}}$ is the eigenvectors of $K$ belonging to the eigenvalues $\pm 1$. On the other hand, for a single qubit with arbitrary state $\varrho$ and orthonormal basis $\{\ket{k}\}$ we have: $C_{l_1}(\varrho;\{\ket{k}\})=2|\braket{k_+|\varrho|k_-}|$. Combining these two equalities we obtain Eq. (\ref{normalized trace-norm asymmetry vs KD coherence and l1-norm coherence}) with the equality replaced by an equality. \qed

Now, consider a set $\Lambda_{\{k\}}$ of Hermitian operators $K$ with a fixed nontrivial spectrum of eigenvalues $\{k\}$. By nontrivial we mean that not all the eigenvalues are equal, so that $K\neq k_0\mathbb{I}$ for some $k_0\in\mathbb{R}$. Then, we obtain the following corollary of Proposition 4. \\
{\bf Corollary 2}.  
The maximum normalized trace-norm asymmetry of a state $\varrho$ relative to the translation groups generated by all $K\in\Lambda_{\{k\}}$ having a fixed nontrivial spectrum $\{k\}$ is bounded from above by the maximum total sum of the absolute imaginary part of the KD quasiprobability associated with $\varrho$ over all possible pairs of the defining orthonormal bases of the Hilbert space: 
\begin{eqnarray}
&&\sup_{K\in\Lambda_{\{k\}}}\tilde{A}_{\rm Tr}(\varrho;K)\nonumber\\
&\le&\sup_{\{\ket{k}\}\in\mathcal{B}_o(\mathcal{H}),\{\ket{x}\}\in\mathcal{B}_o(\mathcal{H})}\sum_{k,x}\big|{\rm Im}{\rm Pr}_{\rm KD}(k,x|\varrho)\big|.
\label{maximum trace-norm asymmetry vs maximum imaginary KD quasiprobability 1}
\end{eqnarray}

Conversely, Eq. (\ref{maximum trace-norm asymmetry vs maximum imaginary KD quasiprobability 1}) can be read as follows. Given a quantum state $\varrho$, the maximum total nonreality of the associated KD quasiprobability over all possible pair of the defining orthonormal bases of the Hilbert space, is bounded from below by the maximum normalized trace-norm asymmetry relative to the translation groups generated by all Hermitian operators $K\in\Lambda_{\{k\}}$. From this viewpoint, and noting the fact that the right-hand side of Eq. (\ref{maximum trace-norm asymmetry vs maximum imaginary KD quasiprobability 1}) is independent of the eigenvalues $\{k\}$ of $K$, the inequality in Eq. (\ref{maximum trace-norm asymmetry vs maximum imaginary KD quasiprobability 1}) can be strengthened as follows. \\
{\bf Corollary 3}. The maximum total sum of the imaginary part of the KD quasiprobability associated with $\varrho$ over all possible pairs of the defining orthonormal bases, is never less than the maximum normalized trace-norm asymmetry of the state $\varrho$ relative to the translation groups generated by all bounded Hermitian operators $K$ on the Hilbert space:  
\begin{eqnarray}
&&\sup_{\{\ket{k}\}\in\mathcal{B}_o(\mathcal{H}),\{\ket{x}\}\in\mathcal{B}_o(\mathcal{H})}\sum_{k,x}\big|{\rm Im}{\rm Pr}_{\rm KD}(k,x|\varrho)\big|\nonumber\\
&\ge& \sup_{K\in\mathcal{O}(\mathcal{H})}\tilde{A}_{\rm Tr}(\varrho;K),
\label{maximum imaginary KD quasiprobability vs maximum normalized trace norm asymmetries}
\end{eqnarray} 
where $\mathcal{O}(\mathcal{H})$ is the set of all bounded Hermitian operators on the Hilbert space $\mathcal{H}$.

We show in the Appendix \ref{Trace-norm asymmetry vs nonreality of KD quasiprobability} that for a single qubit with arbitrary state $\varrho$, the generator $K$ of the translation group which reaches the equality in Eq. (\ref{maximum imaginary KD quasiprobability vs maximum normalized trace norm asymmetries}) has the form: $K_*=k_+\ket{k_+}\bra{k_+}+k_-\ket{k_-}\bra{k_-}$ where $\{k_+,k_-\}$ are the real eigenvalues of $K$ corresponding to the eigenvectors $\{\ket{k_+},\ket{k_-}\}$ satisfying $k_+=-k_-$. This is the case e.g. when $K_*=\vec{n}\cdot\vec{\sigma}$, where $\vec{n}$ is a unit vector, and $\vec{\sigma}=(\sigma_x,\sigma_y,\sigma_z)^{\rm T}$ with $\sigma_x$, $\sigma_y$, and $\sigma_z$ Pauli operators, so that $k_+=-k_-=1$. 

These results show that the nonclassical aspect of quantum mechanics captured by the concept of asymmetry relative to a translation group is related to the nonclassicality captured by the imaginary part of the KD quasiprobability whose one of the defining bases is given by the eigenbasis of the generator of the translation group. It thus suggests that the translational asymmetry of a quantum state may be a key quantum ingredient in diverse quantum phenomena where the nonclassicality captured by the anomaous KD quasiprobability has been shown to play important roles, and vice versa.  

We further obtain an upper bound for the trace-norm asymmetry in terms of state purity. \\
{\bf Proposition 5}. The normalized trace-norm asymmetry of a state $\varrho$ on $d$-dimensional Hilbert space relative to a translation group generated by a Hermitian operator $K$ is bounded from above by the purity of the state, i.e., ${\rm Tr}(\varrho^2)$, as
\begin{eqnarray}
\tilde{A}_{\rm Tr}(\varrho;K)\le\sqrt{(d-1)}\big(d{\rm Tr}(\varrho^2)-1\big)^{1/2}. 
\label{trace-norm versus purity}
\end{eqnarray}
{\bf Proof}. Using the relation between the trace-norm asymmetry and the KD quasiprobability of Eq. (\ref{normalized trace-norm asymmetry vs imaginary KD quasiprobability}), we have
\begin{eqnarray}
&&\tilde{A}_{\rm Tr}(\varrho;K)\nonumber\\
\label{trace-norm versus purity step 1}
&\le&\sup_{\{\ket{x}\}\in\mathcal{B}_o(\mathcal{H})}\sum_{k,x}\big|\sum_{k'\neq k}{\rm Im}(\braket{x|k}\braket{k|\varrho|k'}\braket{k'|x})\big|\\
\label{trace-norm versus purity step 1.5}
&\le&\sup_{\{\ket{x}\}\in\mathcal{B}_o(\mathcal{H})}\sum_k\sum_{x,k'\neq k}|\braket{x|k}\braket{k|\varrho|k'}\braket{k'|x}|\nonumber\\
\label{trace-norm versus purity step 2}
&\le&\sum_k\big(\sum_{k'\neq k,x_*}|\braket{x_*|k}\braket{k|\varrho|k'}|^2\sum_{k''\neq k,x'_*}|\braket{k''|x'_*}|^2\big)^{1/2}\\
\label{trace-norm versus purity step 3}
&=&\sum_k\sqrt{d-1}\big(\sum_{k',x_*}|\braket{x_*|k}\braket{k|\varrho|k'}|^2-\braket{k|\varrho|k}^2\big)^{1/2}\\
\label{trace-norm versus purity step 4}
&=&\sum_k\sqrt{d-1}\big(\braket{k|\varrho^2|k}-\braket{k|\varrho|k}^2\big)^{1/2}\\
\label{trace-norm versus purity step 5}
&\le&\sqrt{(d-1)d}\big(\sum_k\braket{k|\varrho^2|k}-\braket{k|\varrho|k}^2\big)^{1/2}\\
\label{trace-norm versus purity step 6}
&=&\sqrt{(d-1)d}\big({\rm Tr}(\varrho^2)-\sum_k\braket{k|\varrho|k}^2\big)^{1/2}. 
\end{eqnarray}
Here, to get Eq. (\ref{trace-norm versus purity step 1}) we have inserted an identity $\sum_{k'}\ket{k'}\bra{k'}=\mathbb{I}$ and noting the fact that the diagonal terms $k=k'$ are real, in Eq. (\ref{trace-norm versus purity step 2}) $\{\ket{x_*}\}$ is a basis which achieves the supremum and we have used the Cauchy-Schwartz inequality, to get Eq. (\ref{trace-norm versus purity step 3}) we have used the completeness relation and completed the sum over $k'$ (to include also the case $k'=k$), to get Eq. (\ref{trace-norm versus purity step 4}) we have used the completeness relation, and to get Eq. (\ref{trace-norm versus purity step 5}) we have again used the Cauchy-Schwartz inequality. Finally noting that $\sum_{k=1}^d\braket{k|\varrho|k}^2\ge 1/d$ in Eq. (\ref{trace-norm versus purity step 6}) we get Eq. (\ref{trace-norm versus purity}) \qed

Notice that for the maximally mixed state, $\varrho_{\rm mm}=\mathbb{I}/d$, the upper bound in Eq. (\ref{trace-norm versus purity}) is indeed vanishing, as desired since translational asymmetry can be seen as a form of coherence. By contrast, for pure states, the upper bound is given by $d-1$. 

\subsection{Lower bounds: maximum average noncommutativity and uncertainty relations}

We first derive a lower bound for the trace-norm asymmetry. \\
{\bf Lemma 1}. Consider a set $\Lambda_{\{x\}}$ of Hermitian operators $X$ with a fixed nontrivial spectrum of eigenvalues $\{x\}$. Then, the trace-norm asymmetry of $\varrho$ relative to a translation group generated by a Hermitian operator $K$ can be bounded from below as 
\begin{eqnarray}
\tilde{A}_{\rm Tr}(\varrho;K)\ge\sup_{X\in\Lambda_{\{x\}}}|{\rm Tr}([\tilde{X},\tilde{K}]\varrho)|/2. 
\label{lower bound for trace-norm asymmetry wrt maximum average noncommutativity 1}
\end{eqnarray}
{\bf Proof}. Using Eqs. (\ref{normalized trace-norm asymmetry}), (\ref{trace-norm asymmetry is equal to maximum total sum of imaginary part of weak value}) and (\ref{maximum average imaginary part of weak value and incompatibility}), we directly have 
\begin{eqnarray}
&&\tilde{A}_{\rm Tr}(\varrho;K)\nonumber\\
&=&\sup_{\{\ket{x}\}\in\mathcal{B}_o(\mathcal{H})}\frac{1}{\|X\|_{\rm max}\|K\|_{\rm max}}\sum_{x}\|X\|_{\rm max}|{\rm Im}{\rm Tr}(\Pi_{x}K\varrho)|\nonumber\\
&\ge&\sup_{\{\ket{x}\}\in\mathcal{B}_o(\mathcal{H})}\frac{1}{\|X\|_{\rm max}\|K\|_{\rm max}}|{\rm Im}{\rm Tr}(XK\varrho)|\nonumber\\
&=& \sup_{\{\ket{x}\}\in\mathcal{B}_o(\mathcal{H})}|{\rm Im}{\rm Tr}(\tilde{X}\tilde{K}\varrho)|\nonumber\\
&=&\sup_{X\in\Lambda_{\{x\}}}|{\rm Tr}([\tilde{X},\tilde{K}]\varrho)|/2, 
\end{eqnarray}
where we have used $X:=\sum_{x}x\Pi_x$. For a single qubit, it can be again checked analytically as shown in the Appendix \ref{Trace norm asymmetry and average noncommutativity}. \qed

Hence, the trace-norm asymmetry of a state $\varrho$ relative to a translation group generated by $K$ is lower bounded by the maximum average noncommutativity between $K$ and any other possible bounded Hermitian operators $X\in\Lambda_{\{x\}}$ whose eigenbasis spans the Hilbert space, divided by their spectral radiuses. Notice that the lower bound takes a form similar to the lower bound of the Kennard-Weyl-Robertson uncertainty relation.  

Furthermore, noting the fact that the left-hand side of the inequality in Eq. (\ref{lower bound for trace-norm asymmetry wrt maximum average noncommutativity 1}) does not depend on the eigenvalues $\{x\}$ of $X$, the inequality can be further tightened as follows.\\
{\bf Corollary 4}. The trace-norm asymmetry of $\varrho$ relative to a translation group generated by a Hermitian operator $K$ can be bounded from below as 
\begin{eqnarray}
\tilde{A}_{\rm Tr}(\varrho;K)\ge\sup_{X\in\mathcal{O}(\mathcal{H})}|{\rm Tr}([\tilde{X},\tilde{K}]\varrho)|/2, 
\label{lower bound for trace-norm asymmetry wrt maximum average noncommutativity stronger}
\end{eqnarray}
where the supremum is taken over the set $\mathcal{O}(\mathcal{H})$ of all bounded Hermitian operators on the Hilbert space $\mathcal{H}$. 

We show in the Appendix \ref{Trace norm asymmetry and average noncommutativity} that for a single qubit, denoting the eigenvalues of $X$ as $\{x\}=\{x_+,x_-\}$, $x_+,x_-\in\mathbb{R}$, the inequality in Eq. (\ref{lower bound for trace-norm asymmetry wrt maximum average noncommutativity stronger}) becomes equality by choosing $x_+=-x_-$. For example, when $x_+=1$, we may take $X_*=\vec{n}\cdot\vec{\sigma}$ where $\vec{n}$ is a unit vector and $\vec{\sigma}=(\sigma_x,\sigma_y,\sigma_z)^{\rm T}$. 

Combining Corollary 3 and 4, and taking the supremum of both sides of Eq. (\ref{lower bound for trace-norm asymmetry wrt maximum average noncommutativity stronger}) over all bounded Hermitian operators $K\in\mathcal{O}(\mathcal{H})$ on the Hilbert space $\mathcal{H}$, we thus obtain the following ordering of quantities: \\
\begin{eqnarray}
&&\sup_{\{\ket{k}\}\in\mathcal{B}_o(\mathcal{H}),\{\ket{x}\}\in\mathcal{B}_o(\mathcal{H})}\sum_{k,x}\big|{\rm Im}({\rm Pr}_{\rm KD}(k,x|\varrho))\big|\nonumber\\
&\ge&\sup_{K\in\mathcal{O}(\mathcal{H})}\tilde{A}_{\rm Tr}(\varrho;K)\ge\sup_{K\in\mathcal{O}(\mathcal{H})}\sup_{X\in\mathcal{O}(\mathcal{H})}|{\rm Tr}([\tilde{X},\tilde{K}]\varrho)|/2. 
\label{ordering of noncommutativity}
\end{eqnarray}

Now, using Lemma 1, we obtain the following proposition.\\
{\bf Proposition 6}. The trace-norm asymmetries of a state $\varrho$ relative to groups of translation generated by Hermitian operators $K$ and $X$ satisfy the following trade-off relation:
\begin{eqnarray}
\tilde{A}_{\rm Tr}(\varrho;K)\tilde{A}_{\rm Tr}(\varrho;X)\ge\frac{1}{4}\big|{\rm Tr}([\tilde{K},\tilde{X}]\varrho)\big|^2. 
\label{uncertainty relation for trace-norm asymmetries}
\end{eqnarray}
{\bf Proof}. First, exchanging the role of $K$ and $X$ in Eq. (\ref{lower bound for trace-norm asymmetry wrt maximum average noncommutativity stronger}) we have 
\begin{eqnarray}
\tilde{A}_{\rm Tr}(\varrho;X)\ge\sup_{K\in\mathcal{O}(\mathcal{H})}|{\rm Tr}([\tilde{X},\tilde{K}]\varrho)|/2,
\label{lower bound for trace-norm asymmetry wrt maximum average noncommutativity 2}
\end{eqnarray}
where the supremum is taken over all bounded Hermitian operators $K\in\mathcal{O}(\mathcal{H})$ on the Hilbert space $\mathcal{H}$. Multiplying Eqs. (\ref{lower bound for trace-norm asymmetry wrt maximum average noncommutativity stronger}) and (\ref{lower bound for trace-norm asymmetry wrt maximum average noncommutativity 2}), we finally obtain
\begin{eqnarray}
&&\tilde{A}_{\rm Tr}(\varrho;K)\tilde{A}_{\rm Tr}(\varrho;X)\nonumber\\
&\ge&|{\rm Tr}([\tilde{X}_*,\tilde{K}]\varrho)||{\rm Tr}([\tilde{X},\tilde{K}_*]\varrho)|/4\nonumber\\
&\ge&|{\rm Tr}([\tilde{X},\tilde{K}]\varrho)|^2/4, 
\label{trade-off relation for trace-norm asymmetry}
\end{eqnarray}
where $X_*$ and $K_*$ are the Hermitian operators which respectively achieve the supremum in Eqs. (\ref{lower bound for trace-norm asymmetry wrt maximum average noncommutativity stronger}) and (\ref{lower bound for trace-norm asymmetry wrt maximum average noncommutativity 2}). \qed

Proposition 6 clarifies the intuition that when the expectation value of the commutator between the Hermitian operators $K$ and $X$ over $\varrho$ is nonvanishing, then the state $\varrho$ must be asymmetric relative to both the translation group generated by $K$ and that generated by $X$. Moreover, the associated trace-norm asymmetries satisfy the trade-off relation of Eq. (\ref{uncertainty relation for trace-norm asymmetries}).  Let us translate this trade-off relation in the language of coherence. Suppose that the lower bound in Eq. (\ref{uncertainty relation for trace-norm asymmetries}) is nonvanishing. Then the state $\varrho$ cannot be commuting with all the eigenprojectors $\{\Pi_x\}$ of $X$ and with all the eigenprojectors $\{\Pi_k\}$ of $K$. This means that the state is coherent relative to both the orthonormal eigenbases $\{\ket{x}\}$ and $\{\ket{k}\}$. Moreover the amount of respective coherence that are quantified by the trace-norm asymmetries satisfy the trade-off relation of Eq. (\ref{uncertainty relation for trace-norm asymmetries}). Finally, recall that $A_{\rm Tr}(\varrho;K)$ can be seen as the genuine quantum part of the uncertainty of the outcomes of the measurement of the observable $K$ when the system is prepared in the state $\varrho$. Eq. (\ref{uncertainty relation for trace-norm asymmetries}) can thus be seen as the trade-off relation between the genuine quantum part of the uncertainty in measurement of two noncommuting observables \cite{Luo's genuine quantum uncertainty1,Luo's genuine quantum uncertainty2,Korzekwa quantum-classical decomposition,Hall quantum-classical decomposition}. Let us mention that a similar trade-off relation is suggested in Ref. \cite{Luo's genuine quantum uncertainty2}, wherein the genuine quantum uncertainty associated with the measurement of $K$ over $\varrho$ is identified by the Wigner-Yanase skew information defined as $I_{\rm WY}(\varrho;K)=-\frac{1}{2}{\rm Tr}([\sqrt{\varrho},K]^2)$ \cite{Wigner-Yanase skew information}. 

Next, combining Eqs. (\ref{trace-norm asymmetry is upper bounded by the quantum Fisher information}) and (\ref{lower bound for trace-norm asymmetry wrt maximum average noncommutativity stronger}), we obtain the following corollary. \\
{\bf Corollary 5}. Consider a setting whereby a parameter $\theta$ is encoded into the quantum state of a system via a translation unitary generated by $K$ as $\varrho_{\theta}=e^{-iK\theta}\varrho e^{iK\theta}$. Then the quantum Fisher information about $\theta$ contained in $\varrho_{\theta}$ is bounded from below as 
\begin{eqnarray}
\tilde{\mathcal{J}}_{\theta}(\varrho;K)^{1/2}\ge\sup_{X\in\mathcal{O}(\mathcal{H})}|{\rm Tr}([\tilde{X},\tilde{K}]\varrho)|,
\label{lower bound for quantum Fisher information in terms average noncommutativity}
\end{eqnarray}
where $\tilde{\mathcal{J}}_{\theta}(\varrho;K)$ is a normalized quantum Fisher information about $\theta$ in $\varrho_{\theta}$ defined as $\tilde{\mathcal{J}}_{\theta}(\varrho;K):=\mathcal{J}_{\theta}(\varrho;K)/\|K\|_{\rm max}^2$.   

The case of a single qubit is discussed in the Appendix \ref{Fisher information vs avarage noncommutativity for a single qubit}, where equality in Eq. (\ref{lower bound for quantum Fisher information in terms average noncommutativity}) is obtained when the spectrum of $X$, i.e., $\{x\}=\{x_+,x_-\}$, $x_+,x_-\in\mathbb{R}$, satisfies $x_-=-x_+$. Corollary 5 shows that the optimal sensitivity of the state $\varrho$ relative to the translation unitary generated by $K$, or equivalently, the optimal sensitivity in the quantum parameter estimation of $\theta$ conjugate to $K$, is lower bounded by the maximum average noncommutativity between $K$ and any other Hermitian operators $X\in\mathcal{O}(\mathcal{H})$ whose eigenbasis spans the Hilbert space $\mathcal{H}$.  

We thus obtain the following result. \\
{\bf Proposition 7}. Consider two Hermitian operators $K$ and $X$, so that they generate unitary imprinting of scalar parameters to the quantum state of the probe in the protocol of quantum parameter estimation, respectively, as $\varrho_{\theta_K}=e^{-iK\theta_K}\varrho e^{iK\theta_K}$ and $\varrho_{\theta_X}=e^{-iX\theta_X}\varrho e^{iX\theta_X}$, where $\theta_K$ is the scalar parameter conjugate to $K$ and $\theta_X$ is to X. Then, the normalized quantum Fisher information about $\theta_K$ in $\varrho_{\theta_K}$ and about $\theta_X$ in $\varrho_{\theta_X}$ satisfy the following trade-off relation:
\begin{eqnarray}
\tilde{\mathcal{J}}_{\theta_K}(\varrho;K)^{1/2}\tilde{\mathcal{J}}_{\theta_X}(\varrho;X)^{1/2}\ge|{\rm Tr}([\tilde{X},\tilde{K}]\varrho)|^2. 
\label{uncertainty relation for quantum Fisher information}
\end{eqnarray} 
{\bf Proof}. Eq. (\ref{uncertainty relation for quantum Fisher information}) can be directly obtained from Eq. (\ref{lower bound for quantum Fisher information in terms average noncommutativity}) by following similarly the proof of Proposition 6. \qed

Since the quantum Fisher information is a monotonic measure of asymmetry as coherence, Eq. (\ref{uncertainty relation for quantum Fisher information}) admits a similar interpretation as the uncertainty relation of Eq. (\ref{uncertainty relation for trace-norm asymmetries}) for trace-norm asymmetry. Moreover, Eq. (\ref{uncertainty relation for quantum Fisher information}) shows that when the quantum expectation value of the noncommutativity between the Hermitian operators $K$ and $X$ over the state $\varrho$ is not vanishing, then the state must be sensitive relative to the translation unitaries generated by $K$ and by $X$, and their sensitivities as quantified by the quantum Fisher information satisfy the trade-off relation (\ref{uncertainty relation for quantum Fisher information}).    

Finally, from Eqs. (\ref{lower bound for trace-norm asymmetry wrt maximum average noncommutativity 2}) and (\ref{lower bound for quantum Fisher information in terms average noncommutativity}) and following again similarly the proof of Proposition 6, we obtain the following result relating the sensitivity of the state relative to the translation unitary generated by $K$ quantified by the quantum Fisher information, and the coherence of the state relative to the eigenbasis of $X$ quantified by the trace-norm asymmetry. \\
{\bf Proposition 8}. The trace-norm asymmetry of $\varrho$ relative to a translation generated by a Hermitian operator $X$, and the quantum Fisher information about $\theta$ in the state $\varrho_{\theta}$ obtained via a unitary imprinting generated by a Hermitian operator $K$, satisfy the following trade-off relation:
\begin{eqnarray}
\tilde{\mathcal{J}}_{\theta}(\varrho;K)^{1/2}\tilde{A}_{\rm Tr}(\varrho;X)\ge|{\rm Tr}([\tilde{X},\tilde{K}]\varrho)|^2/2. 
\label{uncertainty relation for quantum Fisher information and trace-norm asymmetry}
\end{eqnarray} 

As an implication of the Proposition 8 we have the following corollary.\\
{\bf Corollary 6}. Consider a quantum state $\varrho$ and two Hermitian operators $K$ and $X$. Then, the quantum Fisher information about $\theta$ contained in $\varrho_{\theta}$ obtained via a translation unitary generated by $K$, and the $l_1$-norm coherence of the state $\varrho$ relative to the eigenbasis $\{\ket{x}\}$ of $X$, satisfy the following trade-off relation:
\begin{eqnarray}
\tilde{\mathcal{J}}_{\theta}(\varrho;K)^{1/2}C_{l_1}(\varrho;\{\ket{x}\})\ge|{\rm Tr}([\tilde{X},\tilde{K}]\varrho)|^2/2. 
\label{uncertainty relation for quantum Fisher information and l1-norm coherence}
\end{eqnarray} 
{\bf Proof}. The trade-off relation of Eq. (\ref{uncertainty relation for quantum Fisher information and l1-norm coherence}) can be obtained directly by imposing the inequality (\ref{normalized trace-norm asymmetry vs KD coherence and l1-norm coherence}) of Corollary 1 to Eq. (\ref{uncertainty relation for quantum Fisher information and trace-norm asymmetry}). \qed 

\section{Conclusion and Remarks}  

To conclude, we first showed that the trace-norm asymmetry of a state relative to a translation group is equal to the average absolute imaginary part of the weak value of the generator of the translation, maximized over all possible orthonormal bases of the Hilbert space. Hence, the trace-norm asymmetry of an unknown quantum state can be estimated in experiment using a number of methods for measuring the weak value proposed in the literatures, combined with a classical optimization procedure, in the fashion of hybrid quantum-classical variational circuit which should be implementable using the presently available NISQ hardware. It also suggests the physical and statistical interpretation of the trace-norm asymmetry in terms of the interpretations of the imaginary part of the weak value. 

Using the mathematical link between the trace-norm asymmetry and the nonreal weak value we then derived upper bounds for the trace-norm asymmetry relative to a translation group in terms of the quantum uncertainty of the generator of the translation, the quantum Fisher information about a parameter imprinted via the translation unitary, the imaginary part of the corresponding KD quasiprobability, the $l_1$-norm coherence relative to the eigenbasis of the generator of the translation, and the purity of the state. We also obtain a lower bound in terms of the maximum average noncommutativity between the generator of the translation and any other bounded Hermitian operator on the Hilbert space. We then derived trade-off relations for the trace-norm asymmetry and the quantum Fisher information associated with two noncommuting generators of the translation unitary, with a lower bound reminiscent of that for the Kennard-Weyl-Robertson uncertainty relation. 

We hope that by expressing the geometrical trace-norm asymmetry in terms of the operationally well-defined imaginary part of the weak value and KD quasiprobability, it may shed fresh light on the applications of trace-norm asymmetry in a plethora of fields in which the strange weak values and the nonclassical anomalous values of KD quasiprobability have played crucial roles \cite{Lostaglio KD quasiprobability and quantum fluctuation}. Conversely, our results may suggest new insight to promote the concept of strange weak values and the nonclassical values of KD quasiprobability, which have played important roles in quantum foundation, as useful tools to access the nonclassicality captured by the concepts of asymmetry, coherence, nonclassical correlation, and entanglement, which are the key resources for quantum information processing and quantum technology. It is also interesting to extend the present approach to study the asymmetry relative to general quantum channel \cite{Luo state-channel interaction and asymmetry}.  

\begin{acknowledgments}  
This work is partly funded by the Institute for Research and Community Service, Bandung Institute of Technology with the grant number: 2971/IT1.B07.1/TA.00/2021. It is also in part supported by the Indonesia Ministry of Research, Technology, and Higher Education with the grant number: 187/E5/PG.02.00.PT/2022. I would like to thank two anonymous referees for constructive comments and recommendations, and Joel Federicko Sumbowo for useful discussion. 
\end{acknowledgments} 

\appendix 

\section{Some analytical computations for a single qubit\label{Some analytical computations for a single qubit}}

\subsection{The equality of Eq. (\ref{trace-norm asymmetry is equal to maximum total sum of imaginary part of weak value}) for a single qubit \label{Proposition 1 for a single qubit}}

Assume first that the Hermitian generator of the translation group takes the form  $K=k_0\ket{0}\bra{0}+k_1\ket{1}\bra{1}$, $k_0,k_1\in\mathbb{R}$, where $\{\ket{0},\ket{1}\}$ are the eigenvectors of the Pauli $z$-spin operator $\sigma_z$. Then, computing the trace-norm asymmetry, one directly gets 
\begin{eqnarray}
A_{\rm Tr}(\varrho;K)=\|[\varrho,K]\|_1/2=|k_0-k_1||\braket{0|\varrho|1}|.
\label{trace norm for a single qubit}
\end{eqnarray}
On the other hand, to compute $A_{\rm w}(\varrho;K)$ defined in Eq. (\ref{maximum average imaginary part of weak value and incompatibility}), we need to parameterize the whole orthonormal bases $\{\ket{x(\vec{\lambda})}\}\in\mathcal{B}_o(\mathcal{H})$ of the Hilbert space $\mathcal{H}$,  so that varying the parameters $\vec{\lambda}=(\lambda_1,\dots,\lambda_N)^{\rm T}$ over their ranges of values will scan all the orthonormal bases of the Hilbert space over which we make the optimization. For the two-dimensional Hilbert space of interest, let us use the parameterization of the whole orthonormal bases $\{\ket{x}\}=\{\ket{x_+},\ket{x_-}\}\in\mathcal{B}_o(\mathbb{C}^2)$ based on the Bloch sphere as:
\begin{eqnarray}
\ket{x_+(\alpha,\beta)}&:=&\cos\frac{\alpha}{2}\ket{0}+e^{i\beta}\sin\frac{\alpha}{2}\ket{1};\nonumber\\
\ket{x_-(\alpha,\beta)}&:=&\sin\frac{\alpha}{2}\ket{0}-e^{i\beta}\cos\frac{\alpha}{2}\ket{1}, 
\label{complete set of basis in the x-y plane}
\end{eqnarray}
$\alpha\in[0,\pi]$, $\beta\in[0,2\pi)$. Hence, one can scan all the possible orthonormal bases of the two-dimensional Hilbert space by varying the angular parameters $\alpha$ and $\beta$ over their ranges of values. Using this expression for the defining basis in Eq. (\ref{maximum average imaginary part of weak value and incompatibility}), we directly get
\begin{eqnarray}
&&A_{\rm w}(\varrho;K)\nonumber\\
&=&\sup_{\{\ket{x(\alpha,\beta)}\}\in\mathcal{B}_o(\mathbb{C}^2)}\sum_{x=\{x_+,x_-\}}|{\rm Im}\braket{x(\alpha,\beta)|K\varrho|x(\alpha,\beta)}|\nonumber\\
&=&\max_{(\alpha,\beta)\in[0,\pi]\times[0,2\pi)}|k_0-k_1||\braket{0|\varrho|1}||\sin\alpha||\sin(\beta+\phi_{01})|\nonumber\\
&=&|k_0-k_1||\braket{0|\varrho|1}|=A_{\rm Tr}(\varrho;K),
\label{w-asymmetry is equal to trace-norm asymmetry for a single qubit}
\end{eqnarray}
where $\phi_{01}=\arg\braket{0|\varrho|1}$ and the last equality is just Eq. (\ref{trace norm for a single qubit}). Note that the maximum is obtained for the basis of the form (\ref{complete set of basis in the x-y plane}) with $\alpha=\pi/2$ and $\beta=\pi/2-\phi_{01}$. 

The above result can be generalized to arbitrary Hermitian operator generating a translation unitary of the state on two-dimensional Hilbert space: $K=k_+\ket{k_+}\bra{k_+}+k_-\ket{k_-}\bra{k_-}$, with the eigenvalues $k_+,k_-\in\mathbb{R}$, and the corresponding orthonormal eigenvectors $\{\ket{k_+},\ket{k_-}\}$. First, the trace-norm asymmetry can be computed directly to get, noting Eq. (\ref{trace norm for a single qubit}), 
\begin{eqnarray}
A_{\rm Tr}(\varrho;K)=\|[\varrho,K]\|_1/2=|k_+-k_-||\braket{k_+|\varrho|k_-}|.
\label{trace norm for a single qubit general}
\end{eqnarray}
Let us show that $A_{\rm w}(\varrho;K)$ defined in Eq. (\ref{maximum average imaginary part of weak value and incompatibility}) also yields the same value in accord with Proposition 1. To do this, we first show that for arbitrary state $\varrho$ and Hermitian operator $K$ on finite-dimensional Hilbert space, $A_{\rm w}(\varrho;K)$ is unitarily covariant. Namely, for any unitary transformation $V$, we have  
\begin{eqnarray}
&&A_{\rm w}(V\varrho V^{\dagger};VKV^{\dagger})\nonumber\\
&=&\sup_{\{\ket{x}\}\in\mathcal{B}_o(\mathcal{H})}\sum_x\big|{\rm Im}\braket{x|VK V^{\dagger}V\varrho V^{\dagger}|x}\big|\nonumber\\
\label{proof of the unitary covariant property step 3}
&=&\sup_{\{\ket{x'}\}\in\mathcal{B}_o(\mathcal{H})}\sum_{x'}\big|{\rm Im}\braket{x'|K\varrho|x'}\big|\\
\label{proof of the unitary covariant property step 4}
&=&A_{\rm w}(\varrho;K), 
\end{eqnarray}
where we have defined a new orthonormal basis $\{\ket{x'}\}=\{V^{\dagger}\ket{x}\}$ to get Eq. (\ref{proof of the unitary covariant property step 3}), and Eq. (\ref{proof of the unitary covariant property step 4}) holds since the set of the new orthonormal bases $\{\ket{x'}\}$ is the same as the set of the old orthonormal bases $\{\ket{x}\}$ given by set $\mathcal{B}_o(\mathcal{H})$ of all the orthonormal bases of the same Hilbert space $\mathcal{H}$, so that $\sup_{\{\ket{x}\}\in\mathcal{B}_o(\mathcal{H})}(\cdot)=\sup_{\{\ket{x'}\}\in\mathcal{B}_o(\mathcal{H})}(\cdot)$. 

Now, for the case of a single qubit of interest, let us choose the following unitary transformation: $V=\ket{0}\bra{k_+}+\ket{1}\bra{k_-}$ so that we have $VKV^{\dagger}=k_+\ket{0}\bra{0}+k_-\ket{1}\bra{1}$, and $V\varrho V^{\dagger}=\braket{k_+|\varrho|k_+}\ket{0}\bra{0}+\braket{k_+|\varrho|k_-}\ket{0}\bra{1}+\braket{k_-|\varrho|k_+}\ket{1}\bra{0}+\braket{k_-|\varrho|k_-}\ket{1}\bra{1}$. Noting these facts and using Eq. (\ref{w-asymmetry is equal to trace-norm asymmetry for a single qubit}), we thus obtain 
\begin{eqnarray}
A_{\rm w}(\varrho;K)&=&A_{\rm w}(V\varrho V^{\dagger};VKV^{\dagger})\nonumber\\
&=&|k_+-k_-||\braket{k_+|\varrho|k_-}|=A_{\rm Tr}(\varrho;K), 
\label{TC w-coherence is equal trace-norm asymmetry for a single qubit}
\end{eqnarray} 
where the last equality is just Eq. (\ref{trace norm for a single qubit general}).

\subsection{Trace-norm asymmetry vs quantum standard deviation of Eq. (\ref{trace-norm asymmetry is upper bounded by quantum uncertainty}) for a single qubit \label{trace-norm asymmetry vs quantum uncertainty for a single qubit}}

Assume first, as in the Appendix \ref{Proposition 1 for a single qubit}, the following form of generator of translation group: $K=k_0\ket{0}\bra{0}+k_1\ket{1}\bra{1}$, $k_0,k_1\in\mathbb{R}$. For our purpose, it is convenient to write the state as $\varrho=(\mathbb{I}+r_x\sigma_x+r_y\sigma_y+r_z\sigma_z)/2$, where $(r_x,r_y,r_z)$ are real numbers satisfying $r_x^2+r_y^2+r_z^2=r^2\le 1$, and $(\sigma_x,\sigma_y,\sigma_z)$ are the three Pauli operators. Then one directly obtains 
\begin{eqnarray}
\Delta_K(\varrho)&=&\frac{1}{2}|k_0-k_1|\sqrt{(1-r_z^2)}\nonumber\\
&\ge&\frac{1}{2}|k_0-k_1|\sqrt{(r^2-r_z^2)}\nonumber\\
&=&\frac{1}{2}|k_0-k_1||r_x-ir_y|\nonumber\\
&=&|k_0-k_1||\braket{0|\varrho|1}|\nonumber\\
&=&A_{\rm Tr}(\varrho;K), 
\end{eqnarray}
where the last equality is just Eq. (\ref{trace norm for a single qubit}). Equality is reached when $r=1$, i.e., for pure states as expected. The above result can be generalized to arbitrary Hermitian operator on two-dimensional Hilbert space $K=k_+\ket{k_+}\bra{k_+}+k_-\ket{k_-}\bra{k_-}$, $k_+,k_-\in\mathbb{R}$ and arbitrary state $\varrho$, by first noting that $\Delta_K(\varrho)$, like $A_{\rm Tr}(\varrho;K)$, is unitarily covariant, i.e., $\Delta_{VKV^{\dagger}}(V\varrho V^{\dagger})=\Delta_K(\varrho)$ for arbitrary unitary transformation $V$, and by choosing a unitary transformation $V=\ket{0}\bra{k_+}+\ket{1}\bra{k_-}$, and noting further the fact that the unitary transformation conserves the state purity.  

\subsection{Trace-norm asymmetry vs quantum Fisher information of Eq. (\ref{trace-norm asymmetry is upper bounded by the quantum Fisher information}) for a single qubit\label{Trace-norm asymmetry vs Fisher information for a single qubit}}

Let us write the density operator in terms of its spectral decomposition: $\varrho=\lambda_1\ket{\lambda_1}\bra{\lambda_1}+\lambda_2\ket{\lambda_2}\bra{\lambda_2}$, $\lambda_1,\lambda_2\in\mathbb{R}^+$. Then, assuming $\varrho_{\theta}$ is obtained via a unitary imprinting generated by $K$, i.e., $\varrho_{\theta}=U_{K,\theta}\varrho U_{K,\theta}^{\dagger}$, one has \cite{Braunstein estimation-based UR 2}
\begin{eqnarray}
&&\mathcal{J}_{\theta}(\varrho;K)=\mathcal{J}_{\theta}(\varrho_{\theta};K)\nonumber\\
&=&4\frac{|\lambda_1-\lambda_2|^2}{\lambda_1+\lambda_2}|\braket{\lambda_1|U^{\dagger}_{K,\theta}KU_{K,\theta}|\lambda_2}|^2\nonumber\\
&=&4|\lambda_1-\lambda_2|^2|\braket{\lambda_1|K|\lambda_2}|^2,
\label{quantum Fisher information for a single qubit}
\end{eqnarray}
where we have used the fact that $\lambda_1+\lambda_2=1$ and $U^{\dagger}_{K,\theta}KU_{K,\theta}=K$. On the other hand, one can directly compute the trace-norm asymmetry of $\varrho$ relative to the translation group generated by $K$ in the basis $\{\ket{\lambda_1},\ket{\lambda_2}\}$ to get, 
\begin{eqnarray}
4A_{\rm Tr}(K;\varrho)^2&=&\|[\varrho,K]\|_1^2\nonumber\\
&=&4|\lambda_1-\lambda_2|^2|\braket{\lambda_1|K|\lambda_2}|^2\nonumber\\
&=&\mathcal{J}_{\theta}(\varrho;K), 
\end{eqnarray}
where the last equality is just Eq. (\ref{quantum Fisher information for a single qubit}). Hence, the inequality in Eq. (\ref{trace-norm asymmetry is upper bounded by the quantum Fisher information}) is saturated for arbitrary state of a single qubit.

\subsection{Normalized trace-norm asymmetry vs maximum nonreality of KD quasiprobability of Eq. (\ref{normalized trace-norm asymmetry vs imaginary KD quasiprobability}) for a single qubit\label{Trace-norm asymmetry vs nonreality of KD quasiprobability}}

The Hermitian generator of the translation group can be in general written as $K=k_+\ket{k_+}\bra{k_+}+k_-\ket{k_-}\bra{k_-}$, $k_+,k_-\in\mathbb{R}$. Assume without loss of generality $|k_+|>|k_-|$, so that $\|K\|_{\rm max}=|k_+|$. Then, noting Eq. (\ref{trace norm for a single qubit general}), we have 
\begin{eqnarray}
\tilde{A}_{\rm Tr}(\varrho;K)&=&\frac{|k_+-k_-|}{\|K\|_{\rm max}}|\braket{k_+|\varrho|k_-}|\nonumber\\
&=&\frac{|k_+-k_-|}{|k_+|}|\braket{k_+|\varrho|k_-}|. 
\label{normalized trace-norm asymmetry for a single qubit}
\end{eqnarray}
On the other hand, for a single qubit, we have 
\begin{eqnarray}
\sup_{\{\ket{x}\}\in\mathcal{B}_o(\mathbb{C}^2)}\sum_{k,x}\big|{\rm Im}{\rm Pr}_{\rm KD}(k,x|\varrho)\big|=2|\braket{k_+|\varrho|k_-}|. 
\label{KD coherence for a single qubit}
\end{eqnarray}
See Ref. \cite{Agung KD-nonreality coherence} for a proof. Noting that in this case we also have 
\begin{eqnarray}
\frac{|k_+-k_-|}{|k_+|}\le 2,
\label{fundamental inequality}
\end{eqnarray} 
Eqs. (\ref{normalized trace-norm asymmetry for a single qubit}) and (\ref{KD coherence for a single qubit}) satisfy the inequality of Eq. (\ref{normalized trace-norm asymmetry vs imaginary KD quasiprobability}) of Proposition 4:
\begin{eqnarray}
\tilde{A}_{\rm Tr}(\varrho;K)\le\sup_{\{\ket{x}\}\in\mathcal{B}_o(\mathbb{C}^2)}\sum_{k,x}\big|{\rm Im}({\rm Pr}_{\rm KD}(k,x|\varrho))\big|.
\label{normalized trace-norm asymmetry vs imaginary KD quasiprobability: appendix}
\end{eqnarray}
Next, notice that the equality in Eq. (\ref{fundamental inequality}) and thus the equality in Eq. (\ref{normalized trace-norm asymmetry vs imaginary KD quasiprobability: appendix}) are attained when $|k_+-k_-|=2|k_+|$, which is the case  when $k_-=-k_+$. For example, assume that $|k_+|=1$. Then, we may take $K=\vec{n}\cdot\vec{\sigma}$, where $\vec{n}$ is a unit vector and $\vec{\sigma}=(\sigma_x,\sigma_y,\sigma_z)^{\rm T}$ is the vector of the three Pauli operators. 

\subsection{Normalized trace-norm asymmetry vs maximum average noncommutativity of Eq. (\ref{lower bound for trace-norm asymmetry wrt maximum average noncommutativity 1}) for a single qubit \label{Trace norm asymmetry and average noncommutativity}}

Without loss of generality, we can assume that the generator of the translation group in the case of a single qubit has the following spectral decomposition: $K=k_0\ket{0}\bra{0}+k_1\ket{1}\bra{1}$, $k_0,k_1\in\mathbb{R}$. Now, let us denote the eigenvalues of $X$ as $\{x\}=\{x_+,x_-\}$, $x_+,x_-\in\mathbb{R}$, so that 
\begin{eqnarray}
&&X(\alpha,\beta)\nonumber\\
&=&x_+\ket{x_+(\alpha,\beta)}\bra{x_+(\alpha,\beta)}+x_-\ket{x_-(\alpha,\beta)}\bra{x_-(\alpha,\beta)},\nonumber\\
\label{general Hermitian operator on 2-dimensional Hilbert space}
\end{eqnarray} 
where the eigenvectors $\{\ket{x_+(\alpha,\beta)},\ket{x_-(\alpha,\beta)}\}$ are expressed using the Bloch sphere parameterization as in Eq. (\ref{complete set of basis in the x-y plane}). Furthermore, assume that $|k_0|>|k_1|$ and $|x_+|>|x_-|$, so that $\|K\|_{\max}=|k_0|$ and $\|X\|_{\rm max}=|x_+|$. Then, computing the average noncommutativity between $\tilde{K}$ and $\tilde{X}(\alpha,\beta)$ over the state $\varrho$, and taking the supremum over $(\alpha,\beta)\in[0,\pi]\times[0,2\pi)$, we obtain 
\begin{eqnarray}
&&\sup_{X(\alpha,\beta)\in\Lambda_{\{x_0,x_1\}}}|{\rm Tr}([\tilde{X}(\alpha,\beta),\tilde{K}]\varrho)|/2\nonumber\\
&=&\max_{(\alpha,\beta)\in[0,\pi]\times[0,2\pi)}\frac{|x_+-x_-|}{2|x_+|}\frac{|k_0-k_1|}{|k_0|}|\braket{1|\varrho|0}|\nonumber\\
&\times&|\sin\alpha\sin(\beta+\phi_{01})|\nonumber\\
&=&\frac{|x_+-x_-|}{2|x_+|}\frac{|k_0-k_1|}{|k_0|}|\braket{0|\varrho|1}|\nonumber\\
&\le&\frac{|k_0-k_1|}{|k_0|}|\braket{0|\varrho|1}|\nonumber\\
&=&\tilde{A}_{\rm Tr}(\varrho;K), 
\label{trace-norm asymmetry vs average noncommutativity for a single qubit}
\end{eqnarray} 
in accord with Eq. (\ref{lower bound for trace-norm asymmetry wrt maximum average noncommutativity 1}) of Lemma 1. Here, $\phi_{01}=\arg\braket{0|\varrho|1}$, the maximum is obtained for $X(\alpha,\beta)$ having the form of (\ref{general Hermitian operator on 2-dimensional Hilbert space}) with $\alpha=\pi/2$ and $\beta=\pi/2-\phi_{01}$, the inequality is due to the fact that $|x_+-x_-|/2|x_+|\le 1$, and the last equality is just Eq. (\ref{trace norm for a single qubit general}). Again, the equality in Eq. (\ref{trace-norm asymmetry vs average noncommutativity for a single qubit}) is attained when $x_-=-x_+$. 

\subsection{Normalized quantum Fisher information vs maximum average noncommutativity of Eq. (\ref{lower bound for quantum Fisher information in terms average noncommutativity}) for a single qubit\label{Fisher information vs avarage noncommutativity for a single qubit}}

Writing the density operator in terms of its spectral decomposition, i.e., $\varrho=\lambda_1\ket{\lambda_1}\bra{\lambda_1}+\lambda_2\ket{\lambda_2}\bra{\lambda_2}$, $\lambda_1,\lambda_2\in\mathbb{R}^+$, $\lambda_1+\lambda_2=1$, we have 
\begin{eqnarray}
\tilde{\mathcal{J}}_{\theta}(\varrho,K)&=&\mathcal{J}_{\theta}(\varrho;K)/\|K\|_{\rm max}^2\nonumber\\
&=&4|\lambda_1-\lambda_2|^2|\braket{\lambda_1|K|\lambda_2}|^2/\|K\|_{\rm max}^2,
\end{eqnarray}
where we have used Eq. (\ref{quantum Fisher information for a single qubit}). For the case of a single qubit, let us express the Hermitian operator $X(\alpha,\beta)$ as in Eq. (\ref{general Hermitian operator on 2-dimensional Hilbert space}), where $\alpha\in[0,\pi]$, $\beta\in[0,2\pi)$ are the angular parameters of the Bloch sphere. Then, the maximum average noncommutativity between $\tilde{X}(\alpha,\beta)$ and $\tilde{K}$ in $\varrho$ over $(\alpha,\beta)\in[0,\pi]\times[0,2\pi)$ on the right-hand side of Eq. (\ref{lower bound for quantum Fisher information in terms average noncommutativity}) can be computed directly, in the basis given by the eigenvectors of $\varrho$, to obtain
\begin{eqnarray}
&&\sup_{X(\alpha,\beta)\in\Lambda_{\{x\}}}|{\rm Tr}([\tilde{X},\tilde{K}]\varrho)|\nonumber\\
&=&\max_{(\alpha,\beta)\in[0,\pi]\times[0,2\pi)}\frac{|x_+-x_-|}{|x_+|}|\lambda_2-\lambda_1|\frac{|\braket{\lambda_2|K|\lambda_1}|}{\|K\|_{\rm max}}\nonumber\\
&\times&|\sin\alpha||\sin(\beta-\varphi_{12})|\nonumber\\
&=&\frac{|x_+-x_-|}{|x_+|}|\lambda_2-\lambda_1||\braket{\lambda_2|K|\lambda_1}|/\|K\|_{\rm max}\nonumber\\
&\le&2|\lambda_2-\lambda_1||\braket{\lambda_2|K|\lambda_1}|/\|K\|_{\rm max}\nonumber\\
&=&\sqrt{\tilde{\mathcal{J}}_{\theta}(\varrho,K)}. 
\label{Fisher information vs average noncommutativity - appendix}
\end{eqnarray}
Here, without loss of generality, we have assumed $\|X\|_{\rm max}=|x_+|$, $\varphi_{12}=\arg\braket{\lambda_1|K|\lambda_2}$, and the inequality is due to $|x_+-x_-|/|x_+|\le 2$. Notice that equality in Eq. (\ref{Fisher information vs average noncommutativity - appendix}) is again attained when $x_+=-x_-$. 

\section{Proof of Eq. (\ref{TC w-coherence is upper bounded by quantum standard deviation})\label{maximum total sum of imaginary part is upper bounded by quantum uncertainty}}

First, from the definition of Eq. (\ref{maximum average imaginary part of weak value and incompatibility}) we have 
\begin{eqnarray}
&&A_{\rm w}(\varrho;K)\nonumber\\
&=&\sup_{\{\ket{x}\}\in\mathcal{B}_o(\mathcal{H})}\sum_x\big|{\rm Im}K_{\rm w}(\Pi_x|\varrho)\big|{\rm Tr}(\Pi_x\varrho)\nonumber\\
&\le&\sup_{\{\ket{x}\}\in\mathcal{B}_o(\mathcal{H})}\big(\sum_x({\rm Im}K_{\rm w}(\Pi_x|\varrho))^2{\rm Tr}(\Pi_x\varrho)\big)^{1/2}, 
\label{TI w-coherence and MSE of estimation of generator}
\end{eqnarray}
where we have made use of the Jensen inequality. Next, noting that $({\rm Im}K_{\rm w}(\Pi_x|\varrho))^2=|K_{\rm w}(\Pi_x|\varrho)|^2-({\rm Re}K_{\rm w}(\Pi_x|\varrho))^2$, and inserting into Eq. (\ref{TI w-coherence and MSE of estimation of generator}), we get
\begin{eqnarray}
&&A_{\rm w}(\varrho;K)\nonumber\\
\label{from weak measurement to quantum uncertainty 0}
&\le&\Big(\sum_{x_*}\Big(\Big|\frac{{\rm Tr}(\Pi_{x_*}K\varrho)}{{\rm Tr}(\Pi_{x_*}\varrho)}\Big|^2-{\rm Re}\Big(\frac{{\rm Tr}(\Pi_{x_*}K\varrho)}{{\rm Tr}(\Pi_{x_*}\varrho)}\Big)^2\Big)\nonumber\\
&\times&{\rm Tr}(\Pi_{x_*}\varrho)\Big)^{1/2}\\
\label{from weak measurement to quantum uncertainty 1}
&\le&\Big(\sum_{x_*}\frac{|{\rm Tr}(\Pi_{x_*}K\varrho )|^2}{{\rm Tr}(\Pi_{x_*}\varrho)}-\big(\sum_{x_*}{\rm Re}{\rm Tr}(\Pi_{x_*}K\varrho)\big)^2\Big)^{1/2},
\end{eqnarray}
where we have used the definition of the weak value of Eq. (\ref{complex weak value}) to get Eq. (\ref{from weak measurement to quantum uncertainty 0}), $\{\ket{x_*}\}$ is a basis which achieves the supremum, and to get Eq. (\ref{from weak measurement to quantum uncertainty 1}) we have applied the Jensen inequality, i.e., $(\sum_{x_*}{\rm Re}{\rm Tr}(\Pi_{x_*}K\varrho))^2=(\sum_{x_*}\frac{{\rm Re}{\rm Tr}(\Pi_{x_*}K\varrho)}{{\rm Tr}(\Pi_{x^*}\varrho)}{\rm Tr}(\Pi_{x^*}\varrho))^2\le\sum_{x_*}(\frac{{\rm Re}{\rm Tr}(\Pi_{x_*}K\varrho)}{{\rm Tr}(\Pi_{x_*}\varrho)})^2{\rm Tr}(\Pi_{x_*}\varrho)$. Finally, applying the Cauchy-Schwartz inequality to the numerator in the first term on the right-hand side of Eq. (\ref{from weak measurement to quantum uncertainty 1}), i.e., $|{\rm Tr}(\Pi_{x_*}K\varrho)|^2=|{\rm Tr}((\Pi_{x_*}^{1/2}K\varrho ^{1/2})(\varrho ^{1/2}\Pi_{x_*}^{1/2}))|^2\le{\rm Tr}(\Pi_{x_*}K\varrho K){\rm Tr}(\Pi_{x_*}\varrho)$, and using the completeness relation $\sum_{x_*}\Pi_{x_*}=\mathbb{I}$, we obtain
\begin{eqnarray}
A_{\rm w}(\varrho;K)&\le&\big({\rm Tr}(K^2\varrho)-({\rm Tr}(K\varrho))^2)^{1/2}\nonumber\\
&=&\Delta_K(\varrho), 
\end{eqnarray} 
as claimed. 
\qed

\end{document}